\documentclass[journal,transmag]{IEEEtran}
 \usepackage{graphicx}
\usepackage{color}	
\usepackage{subcaption}
\usepackage{caption}
 \usepackage[utf8]{inputenc}

\begin{document}
 
\title{Broadcast Strategies and Performance Evaluation of IEEE 802.15.4 in Wireless Body Area Networks WBAN}

%

 \author{\IEEEauthorblockN{Wafa Badreddine\IEEEauthorrefmark{1},
Claude Chaudet\IEEEauthorrefmark{2},
Federico Petruzzi\IEEEauthorrefmark{1},and
Maria Potop-Butucaru\IEEEauthorrefmark{1}}
\IEEEauthorblockA{\IEEEauthorrefmark{1}UPMC Sorbonne Universites, LIP6-CNRS UMR 7606, France, LastName.FistName@lip6.fr}
\IEEEauthorblockA{\IEEEauthorrefmark{2}Institut Mines-Telecom, Telecom ParisTech, CNRS LTCI UMR 5141, claude.chaudet@telecom-paristech.fr}
\thanks{This work was funded by SMART-BAN project (Labex SMART) http://www.smart-labex.fr. En extended abstract of this work \cite{BCPP15} has been published in ACM MSWIM 2015.}}

\maketitle

\begin{abstract}
The rapid advances in sensors and ultra-low power wireless communication has enabled a new generation of wireless sensor networks: Wireless Body Area Networks (WBAN). To the best of our knowledge the current paper is the first to address broadcast in WBAN. We first analyze several broadcast strategies inspired from the area of Delay Tolerant Networks (DTN). The proposed strategies are evaluated via the OMNET++ simulator that we enriched with realistic human body mobility models and channel models issued from the recent research on biomedical and health informatics. Contrary to the common expectation, our results show that existing research in DTN cannot be transposed without significant modifications in WBANs area. That is, existing broadcast strategies for DTNs do not perform well with human body mobility. However, our extensive simulations give valuable insights and directions for designing efficient broadcast in WBAN.
Furthermore, we propose a novel broadcast strategy that outperforms the existing ones in terms of end-to-end delay, network coverage and energy consumption.
Additionally, we performed investigations of independent interest related to the ability of all the studied strategies to ensure the total order delivery property when stressed with various packet rates. These investigations open new and challenging research directions.
\end{abstract}


 \section{Introduction}
\label{sec:introduction}
Wireless Body Area Networks (WBAN) open an interdisciplinary area within Wireless Sensor Networks (WSN) research, in which sensors are used to monitor, collect and transmit medical signs and other measurements of body parameters. The intelligent sensors can be integrated into clothes (wearable WBANs), or placed directly on or inside a body. If typical applications target personalized, predictive, preventive and participatory healthcare, WBANs also have interesting applications in military, security, sports and gaming fields. Healthcare workers, for instance, are really in demand of systems that permit a continuous monitoring of elderly people or patients to support them in their daily life. WBANs history is just at its beginning, and many news and improvements are expected in near future.

Body area networks differ from typical large-scale wireless sensor networks in many aspects: the size of the network is limited to a dozen of nodes, in-network mobility follows the body movements and the wireless channel has its specificities. Links have, in general, a very short range and a quality that varies with the wearer's posture, but remains low in the general case. Indeed, the transmission power is kept low, which improves devices autonomy and reduces wearers electromagnetic exposition. Consequently, the effects of body absorption, reflections and interference cannot be neglected and it is difficult to maintain a direct link (one-hop) between a data collection point and all WBAN nodes. Multi-hop communication represent a viable alternative, but multi-hop communication protocols proposed in literature are not optimized for this specific mobility pattern. Adapted algorithms could easily take advantage of the particular connection changes pattern to limit control traffic, preserving bandwidth, energy and limiting radio emissions. Delay-Tolerant Networks protocols seem particularly relevant in this context, as they are designed to tackle the intermittent connectivity, thanks to their store-and-forward philosophy. However, they usually either suppose a perfectly known mobility as in interplanetary networks, or a fully random mobility. WBANs mobility exhibit a certain regularity that could be utilized, with a certain degree of randomness.

In this paper, we concentrate on the sole {\em broadcasting} problem. We analyze the behavior of various broadcast strategies adapted from DTN literature and propose an alternative strategy from the analysis of the strengths and weaknesses of the different approaches. We compare network coverage, completion delay and required transmissions of $9$ different algorithms over real human body mobility traces.

The paper is organized as follow: Section \ref{sec:bkg} presents the broadcast problem. Section~\ref{sec:broadcast} introduces relevant related works and presents various broadcast strategies, including an original contribution, that we have compared through simulation in Section \ref{sec:simulation}. Section \ref{sec:simulation} first describes the limitations of the channel models used so far in OMNET++, then describes the realistic channel model we implemented and finally presents and analyzes the evaluation results from various broadcast strategies in terms of end-to-end delay, energy and node coverage.


\section{Background}
\label{sec:bkg}
In the past decade, several routing protocols have been proposed for different types of multi-hop networks (ad-hoc, sensors, vehicular or DTN). Most of these protocols rely on the periodical update of a cost function, the algorithm seeking for the minimal cost paths across the network. The cost function can be based on any network-related parameter (delay, number of hops, congestion, stability, QoS parameters, etc.), system-related parameter (e.g. battery level, temperature or available memory), or application level (e.g. security or measured data). The algorithm can rely on a local vision of the network, each node selects locally the path that exhibits the smallest cost to reach the destination, or it computes following a global optimization process that tries to minimize the total, network-wide, cost. Multiple metrics can be combined, generally through a linear combination (weighted sum).

In most cases, in-network mobility is handled by updating frequently the measured parameters and the cost, and by selecting dynamically the best path. Human body mobility, however, has a quasi-periodical behavior that could be exploited. Links appear and disappear quickly, in a quasi-regular pattern and learning this pattern, its regularities and its deviations, could influence the protocol design, reducing the frequency of cost updates. Such patterns were already utilized in the area of interplanetary networks (IPN), which leaded to standard proposals in the DTN field. However, on-body wireless channel is far less reliable and less predictable than the satellite-to-satellite link and mobility has a periodical component but is never perfectly regular. Therefore even though DTN paradigms are clearly of interest, adaptations are necessary.

\section{Broadcast Strategies}
\label{sec:broadcast}
Based on the amount of knowledge that is available, the broadcast algorithms for DTN are divided in two big families \cite{DTN1}: dissemination (a.k.a. flooding) algorithms and forwarding (a.k.a. knowledge) algorithms. Flooding consists in disseminating every message to every node in the network. This strategy is adopted when no knowledge on the networks is available. Flooding maximizes the probability that a message is successfully transferred to its destination(s) by replicating multiple copies of the message, hoping that one will succeed in reaching its destination.

While providing near-optimal performance regarding delay, the primary limitation of these protocols is their energy consumption and their overhead due to excessive packet transmissions. Blind flooding algorithms generate numerous useless transmissions, while more intelligent algorithms usually rely on the knowledge of the nodes movement to predict spatio-temporal connectivity. The more precisely the mobility pattern is characterized, the more optimized the dissemination will be. However, acquiring this information also has a cost and there is a subtle balance to find between duplicate data packets and control messages.

\subsection{Related works}

Optimizing flooding in wireless multihop networks has been a constant concern over the past decades. In their seminal paper \cite{Ni:1999sd}, Ni {\em et al.} the {\em broadcast storm} problem in mobile ad-hoc networks: blind flooding generates numerous transmissions all over the network in the same time frame, causing collisions, increased contention and redundant transmissions. Ni {\em et al.} analyze different strategies to alleviate this effect. Each node can simply condition its retransmission to a constant probability, or take a forwarding decision based on the number of copies received during a certain time frame, on the distance between the source and the destination or on the location of the nodes if it is available. They also examine the effect of partitioning the network in 1-hops clusters, the cluster heads forming a dominating set among the network. The dominating sets-based approaches will be the source of numerous contributions, but is not really relevant in WBANs, as the size and diameter of the network remains very limited.

Many works have explored the probabilistic flooding approach, starting from Haas {\em et al.} \cite{Haas:2002iv}, who identify the existence of a threshold on the forwarding probability below which probabilistic flooding fails. Sasson {\em et al.}, in \cite{Sasson:2003hr} will push the analysis further using percolation theory to find the threshold that triggers such phase transition on random graphs. These works are not directly applicable to WBANs, as the mobility pattern is very different and non-homogeneous, but they show that the networks shape and dynamics have an influence on the optimal forwarding probability.

However, even with a probabilistic transmission, a node may receive multiple instances of the same message from various paths. Vahdat {\em et al.}, in \cite{Vahdat:epidemic}, introduce {\em epidemic routing}. They suppose that a MANET is formed by multiple mobile clusters that eventually meet. Inside a cluster, they disseminate information through basic flooding and when two nodes belonging to different clusters meet, they exchange a {\em summary vector} containing the messages ID they already possess. They then exchange only the missing messages that will reach new clusters this way. Considering that the nodes were divided in connected groups without any guarantee from global connectivity was a first step towards delay-tolerant networks.

When connection is intermittent, unicast routing towards a single destination cannot be based simply on a classical shortest path algorithm like link-state routing. Indeed, the time required to collect all links information is too high compared to the links intermittency. That's why several papers such as \cite{Merugu:2004it,Jain:2004uj} propose and evaluate mobility-aware shortest path algorithms. Besides the chosen strategies, these contributions confirmed that the transmissions efficiency increases with the knowledge of the mobility pattern. In other words, knowing how nodes move allows to reduce the number of unnecessary transmissions and the delivery delay.

In \cite{Harras:2005qi}, the authors compared various strategies for controlling flooding in delay tolerant networks. They compare basic probabilities approaches with time-to-live approaches in terms of number of hops or time stamps and with explicit notification when the destination has received the message.
Further more, \cite{spray} proposes an interesting flooding-forwarding strategy, \textbf{Spray and Wait}.
The spray phase (flooding) is as follows: for every message originating at a source node, copies are initially spread and forwarded by the source and possibly other nodes receiving a copy to distinct relays. In the wait phase (forwarding), if the destination is not found in the spraying phase, each node carrying a message copy performs a direct transmission (will forward the message only to its destination).
 A variant of this algorithm is \textit{Spray and Focus}. The forwarding requirement is to have some knowledge of the network, called in this article as "utility function" that represent how much the node is useful to reach the destination.
The authors of \cite{EBP} propose another strategy to optimize broadcasting in DTN: the {\em k-neighbor} broadcast scheme, in which a packet is retransmitted if and only if the number of neighbors present exceed a threshold, $K$, and if at least one of these neighbors did not receive the message yet. Even if the implementation is different, we find back here the concepts behind epidemic routing and the notion of not systematically transmitting a packet to reduce the number of transmission. The simulation-based evaluation shows a good performance and study the influence of the threshold on the number of neighbors, $K$. Yet, this threshold is the same all across the network.

In the past decade, several routing protocols have been proposed for WBANs that can be classified with respect to their aims in different categories \cite{WBAN,WBAN1}: temperature based routing protocols, cross layer, cost-effective routing protocols, QoS-based routing protocols. Most of these proposals are not suitable when sensors are external (on the body and not in the body). DTN like solutions \cite{AMR} and many other WBAN proposals (such as EDSR \cite{EDSR}) do not take into account the mobility of the human-body: during the transmissions of the message over the path already computed, disconnections can happen causing failures.

The authors of \cite{Quwaider:2009gh} study the case of unicast routing under the assumption that postural changes provoke network disconnections. They propose a store-and-forward approach and use a probabilistic proactive routing approach, defining stochastic links costs. They show through simulation and experiments that taking into account the particular mobility of the body improves transmission delay when compared to a traditional probabilistic algorithm for DTN. The same authors present in \cite{Quwaider:2009ur} an adaptation of their algorithm taking into account location-aware networks: a node forwards packets if this action results in bringing the data closer (physically) to the destination.

\subsection{Old and new broadcast strategies}
\label{BS}
This excerpt from the literature shows that there is room for optimizing the flooding procedure by taking into account the specific mobility pattern. When a node receives a flooded packet, it needs to take a decision whether to forward it or not, to which of its neighbors and when. However, acquiring the necessary information on the nodes mobility is not free. It either requires a precise nodes localization mechanism, which is typically achieved by measuring round-trip time of flights between couple of nodes in IR-UWB body area networks, or requires to rely on \emph{hello messages} to learn connections/disconnection patterns. In both cases the improvement realized by the mobility pattern characterization could be lost by the necessary control traffic. That's why we wished to compare various strategies with different levels of knowledge, in a scenario that is representative of a real WBAN.

More specifically, we will compare the following strategies:
\begin{itemize}
\item \textit{Flooding}: Nodes retransmit each received packet while its TTL (Time-To-Live) greater than $1$.

\item \textit{Plain flooding}: Nodes retransmit the packet only the first time they receive it. So, each packet will be retransmitted only one time by each node.

\item \textit{Pruned flooding}: Nodes retransmit each packet to only $K$ nodes chosen randomly.

\item \textit{Probabilistic flooding (P=0.5)}: Nodes decide to retransmit packets based on a probability $P$. For each received packet, nodes choose a random variable and compare it with $P$. For simulations, we set this probability to a constant value equal to $0.5$.

\item \textit{Probabilistic flooding (P=P/2)} Nodes decide to retransmit packets based on a probability $P$ too. For each received packet, nodes choose a random variable and compare it with $P$. For this strategy, $P$ decreases as the number of retransmissions increases. The first instance of the packet is retransmitted with a probability equal to $1$ (initially $P=1$, then with probability equal to $0.5$ ($P=P/2$) then $0.25$, etc.

\item \textit{Tabu flooding} For this strategy, extra information is added to the original message indicating for each node in the network if it has already received the message or not. Consequently each node retransmits only to the nodes not yet covered.

\item \textit{EBP (Efficient Broadcast Protocol)}: We adapted this strategy to WBAN context from \cite{EBP}. In this protocol nodes decide to retransmit a packet under two conditions: Only when they are surrounded by at least $K$ neighbors, and if at least one of these neighbors has not received the packet yet. While in \cite{EBP} the threshold value $K$ is fixed and uniform across the network, we adapt its value to the WBAN environment, setting a higher value for the gateway (\emph{sink} and a lower one for the peripheral nodes who are less likely to encounter uncovered nodes. In our implementation the $K$ value is set to $3$ for the gateway (chest), to $1$ for the head, ankle and the wrist (peripheral nodes) and to $2$ for the other nodes. To update neighbors' map, nodes relies on control message (\emph{Hello message}), while updating the information concerning who did or not received the message relies on both control and data messages.

\item \textit{MBP (Mixed Broadcast Protocol)}: We proposed this strategy in \cite{BCPP15} as a mix between the \emph{dissemination-based} and \emph{knowledge-based} approaches. The broadcast begins as a basic flooding algorithm (i.e \emph{Flooding} strategy). When a node receives a message, it checks the number of hops $NH$ this message has traveled since its emission by the \emph{sink}, and compares it with a threshold on the number of hops $\Delta$:
\begin{itemize}
\item As long as $NH < \Delta$, node forwards the packet, using simple flooding, simply transmitting the packet to all neighbors in its range provided that its TTL greater than $1$.
\item When $NH = \Delta$, node waits during a time $T$ to receive up to $Q$ acknowledgments.
\begin{itemize}
\item if it receives a number of acknowledgments greater or equal than $Q$, the node stops rebroadcasting the message.
\item if it fails to receive $Q$ acknowledgments, it simply rebroadcasts the packet.
\end{itemize}
\item When $NH > \Delta$, node applies the same strategy as when $NH = \Delta$, but it sends in addition an acknowledgment to the node it received the packet from.
\end{itemize}

\item \textit{\textbf{OptFlood (Optimized Flooding)}}: Our novel strategy described below that outperforms the strategies adapted from DTN or WSN but also the MBP strategy we proposed in \cite{BCPP15}.
\end{itemize}

The simulation results we detail in Section \ref{sec:simulation} allow us to come out with an interesting set of conclusions:
\begin{itemize}
\item Regarding the average end-to-end delay, i.e. the time required for a message to reach all the nodes, both dissemination (flooding-based approaches) and knowledge-based (EPB-like) protocols spend more time to cover the peripheral nodes (which represent about 20\,\% of the network) compared to the rest of the network.
\item When looking at the volume of control traffic, using an EBP-like protocol makes the number of unnecessary retransmissions increase, due to an increased volume of control packets in the network used mainly to discover neighbors.
\item For \emph{Flooding}, it is very important to set a good $TTL$ value in order to control the retransmissions. However, it ensures a fast network coverage, indeed, it represents the best end-to-end delay for all postures.
\item With \emph{EBP}, using a relatively high threshold on the number of neighbors required to transmit ($K$), works well only if the Hello-interval $I$ is short enough to have a "realistic" representation of the network. Otherwise, with large Hello-interval only the last Hello message received is reliable, while the other neighbors saved in the neighbor-table could be obsolete.
\item Knowledge-based algorithms include control messages such as neighborhood discovery messages. This approach does not work well in the WBAN scenario because it may happen that during the communication (exchange of control messages) to "take decision" the connections disappear. Moreover this increases further transmissions, receptions, and related problems such as collisions and interferences.
 \item For \emph{MBP}, since it mixes flooding-based and knowledge-based approaches, a significant amelioration is noticed. Results show a good percentage of covered nodes (close to \emph{Flooding} strategy) with $50$\%\ less transmissions and receptions comparing to \emph{Flooding}. However, looking at the end-to-end delay, we noticed that although \emph{MBP} shows a better delay than the other strategies, the difference with the end-to-end delay of \emph{Flooding} strategy remains important.
 \end{itemize}
 These conclusions guided us to define a novel protocol \emph{OptFlood (Optimized Flooding)} by taking into account the strengths and weaknesses of the basic strategy \emph{Flooding}. The latter shows far better performances considering the coverage of the network and specially the end-to-end delay. Unfortunately, the important number of transmissions and receptions makes the use of this strategy inadequate for WBAN environment where a low energy consumption is required. \emph{Optimized Flooding} is a revised version of \emph{Flooding} whose purpose is to keep the good end-to-end delay given by \emph{Flooding} while lowering energy consumption with the simplest way and the minimum cost.
 \newline \indent \emph{Optimized Flooding} uses two variables: \textbf{CptLocal} and \textbf{cptGlobal}, described below:
\begin{itemize}
 \item \textbf{cptGlobal} Each message carries \emph{cptGlobal} that can be read and updated by all nodes in the network that receive this message.
 \item \textbf{CptLocal} is a local variable to each node. It can be read and modified by its owner. \emph{cptLocal} is setter to \emph{cptGlobal} of the last received message. In other words, \emph{cptLocal} is the local copy of \emph{cptGlobal}.
\end{itemize}
Our algorithm proceeds as follow: When a node receives a message:
\begin{itemize}
 \item For the first time: the node increments message'\emph{cptGlobal}, updates its \emph{cptLocal} and rebroadcasts the message.
 \item For the $x \geq 2$ time:
 \begin{itemize}
 \item The node verifies if it already incremented the \emph{cptGlobal} in this copy of the message. In case not, the node increments the value of \emph{cptGlobal}.
 \item The node compares \emph{cptGlobal} with \emph{cptMax} which corresponds to the number of nodes in the network and to the maximum value that \emph{cptGlobal} could reach.
 \begin{itemize}
  \item If \emph{cptGlobal} $=$ \emph{cptMax}, this means all nodes already received the message ie $100\%$ of the network is covered so the node stops rebroadcasting this message and all the incoming copies of this message.
 \item If \emph{cptGlobal} $<$ \emph{cptMax}, this means that there are still not covered nodes. Then the node compares the value of \emph{cptGlobal} with the value of \emph{cptLocal}:
 \begin{itemize}
 \item if \emph{cptGlobal} $\leq$ \emph{cptLocal}: which means a message with a \emph{cptGlobal} equal to the current \emph{cptLocal} has already been broadcasted in the neighborhood. This copy of message is obsolete hence the node discards it.
 \item if \emph{cptGlobal} $>$ \emph{cptLocal}: node updates its \emph{cptLocal} with the new value of \emph{cptGlobal} and rebroadcasts the message in the neighborhood with the freshest value of \emph{cptGlobal}.
 \end{itemize}
 \end{itemize}
 \end{itemize}
\end{itemize} 

\section{Simulation settings and results}
\label{sec:simulation}
In order to test the algorithms described above (Section \ref{BS}) in a specific WBAN scenario, we implemented them under the Omnet++ simulator. Omnet++ includes a set of modules that specifically model the lower network layers of WSN and WBAN through the Mixim project~\cite{mixim}. It includes a set of propagation models, electronics and power consumption models and medium access control protocols.
The MoBAN framework for Omnet++ adds mobility models for WBANs of $12$ nodes in $4$ different postures. Unfortunately studying in deep the MoBAN code, we discovered that it mainly focuses on the mobility resulting from the change of position, rather than describing coherent and continuous movements. Besides, it models the movement of each node with respect to the centroid of the body and the signal attenuation between couples of nodes is approximated with a simple propagation formula that is not accurate enough to model low-power on-body transmission. It does not model absorption and reflection effects due to the body, alterations due to the presence of clothes and eventually interference from other technologies at the same frequency since 2.4\,GHz is a crowded band.

\subsection{Channel model}
We therefore decided to implement a realistic channel model published in \cite{channel} over the physical layer implementation provided by the Mixim framework. This channel model of an on-body $2.45\,GHz$ channel between $7$ nodes, that belong to the same WBAN, using small directional antennas modeled as if they were $1.5cm$ away from the body. Nodes are assumed to be attached to the human body on the head, chest, upper arm, wrist, navel, thigh, and ankle. 

\begin{figure}[htbp]
\centering
 \begin{subfigure}{0.9\columnwidth}
 \centering
 \includegraphics[width=\textwidth,height=2.7cm]{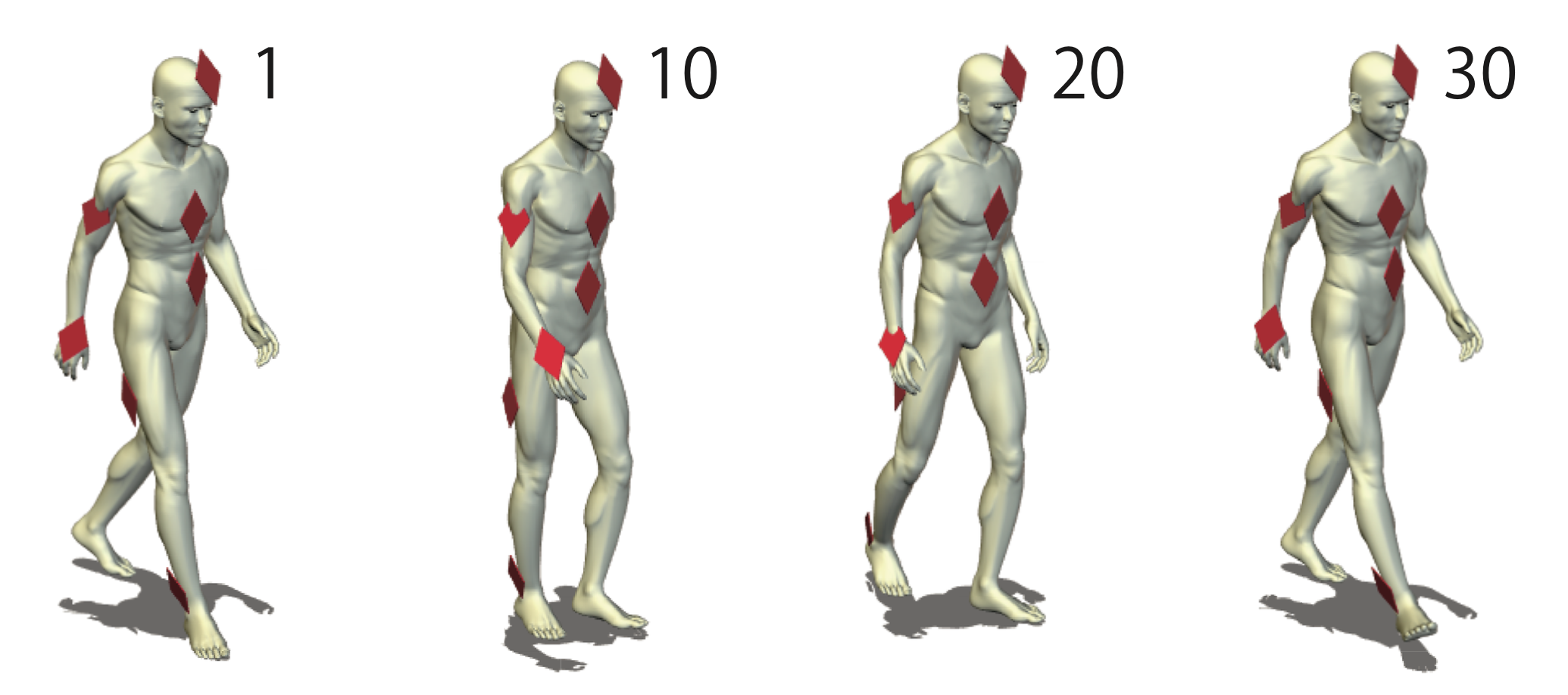}
 \caption{Walking}
 \end{subfigure}
 \begin{subfigure}{0.9\columnwidth}
 \centering
 \includegraphics[width=\textwidth,height=2.5cm]{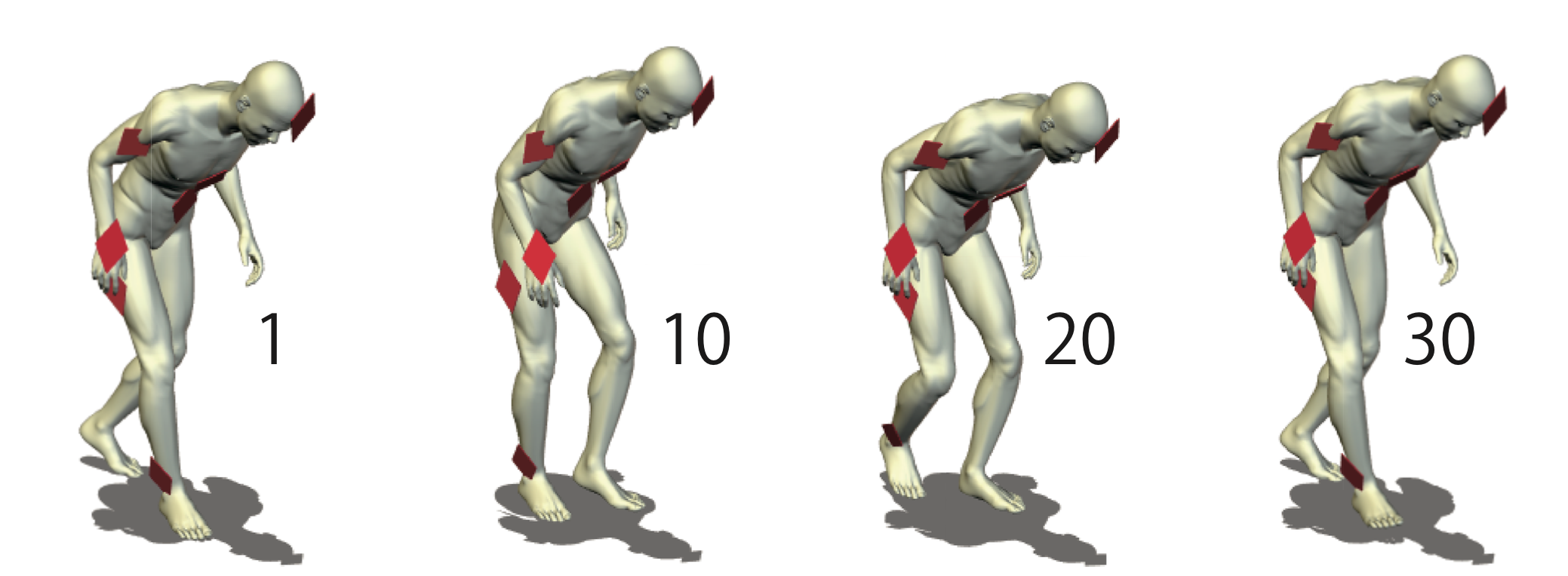}
 \caption{Walking weakly}
 \end{subfigure}
 \begin{subfigure}{0.9\columnwidth}
 \centering
 \includegraphics[width=\textwidth,height=2.5cm]{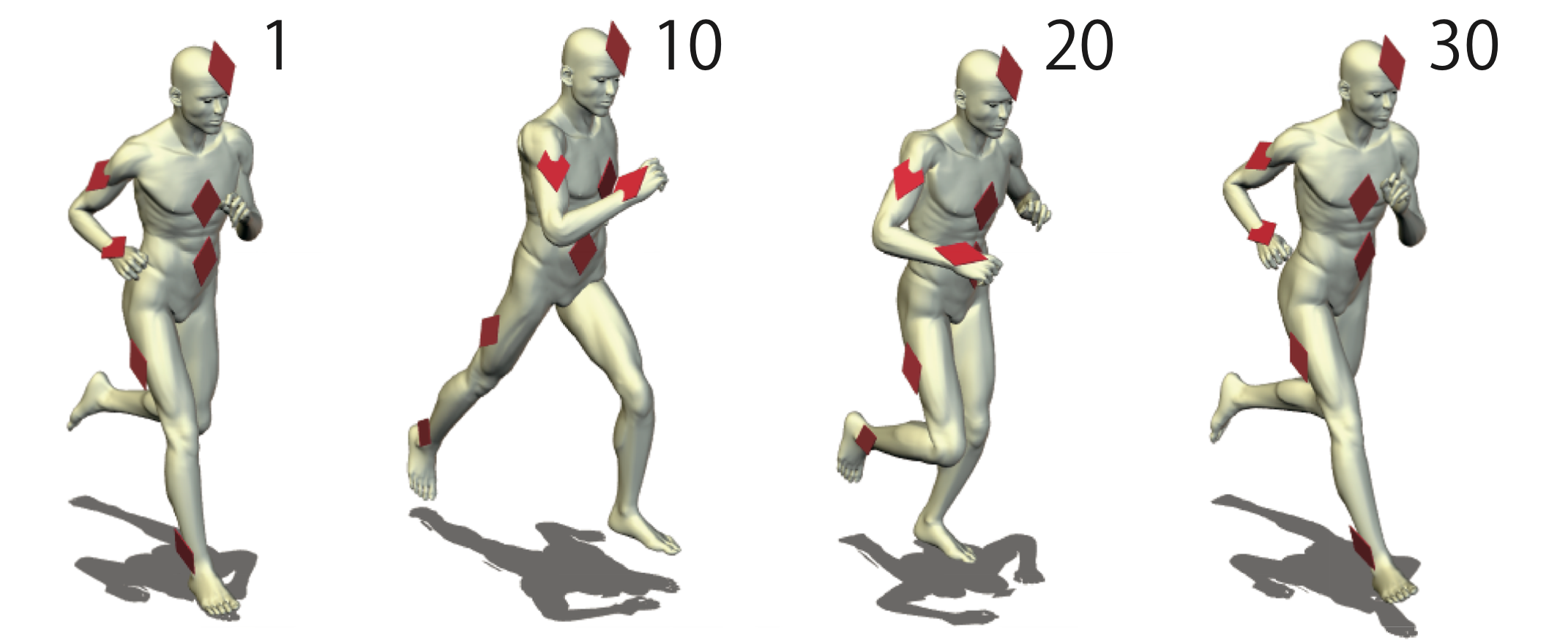}
 \caption{Running}
 \end{subfigure}
 \begin{subfigure}{0.9\columnwidth}
 \centering
 \includegraphics[width=\textwidth,height=2.5cm]{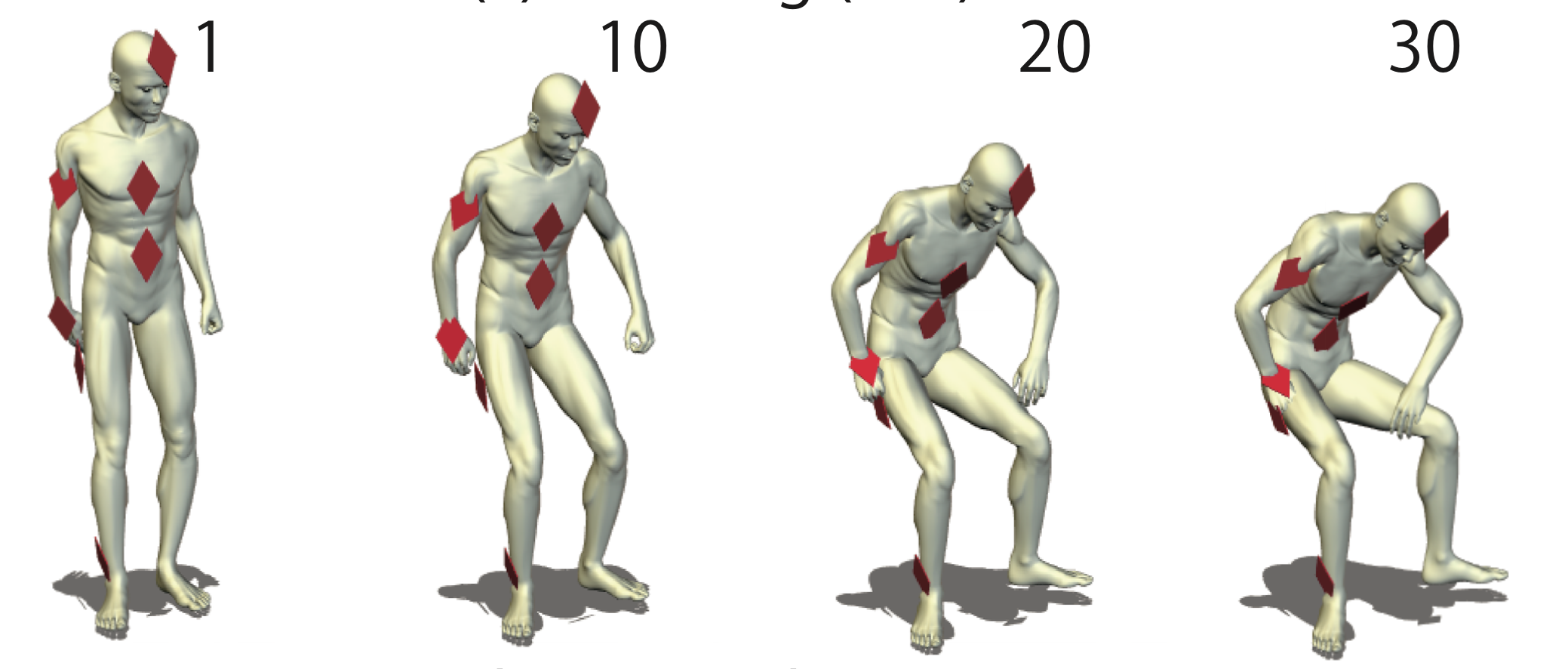}
 \caption{Sitting down}
 \end{subfigure}
 \begin{subfigure}{0.9\columnwidth}
 \centering
 \includegraphics[width=\textwidth,height=2.5cm]{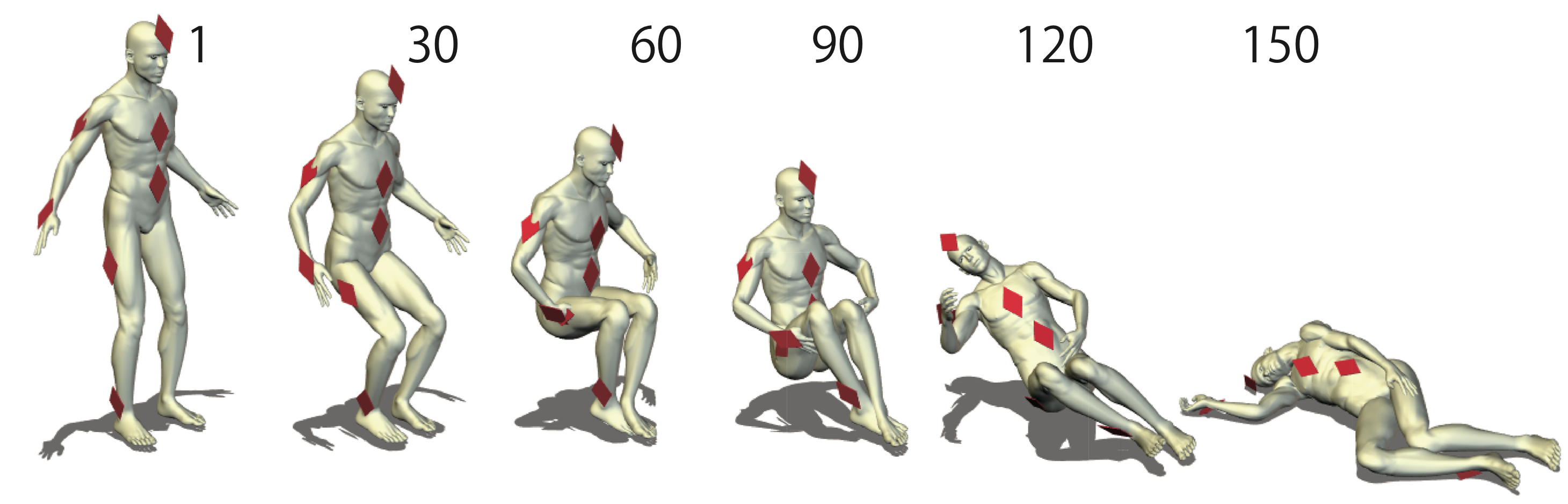}
 \caption{Lying down}
 \end{subfigure}
 \begin{subfigure}{0.9\columnwidth}
 \centering
 \includegraphics[width=\textwidth,height=2.5cm]{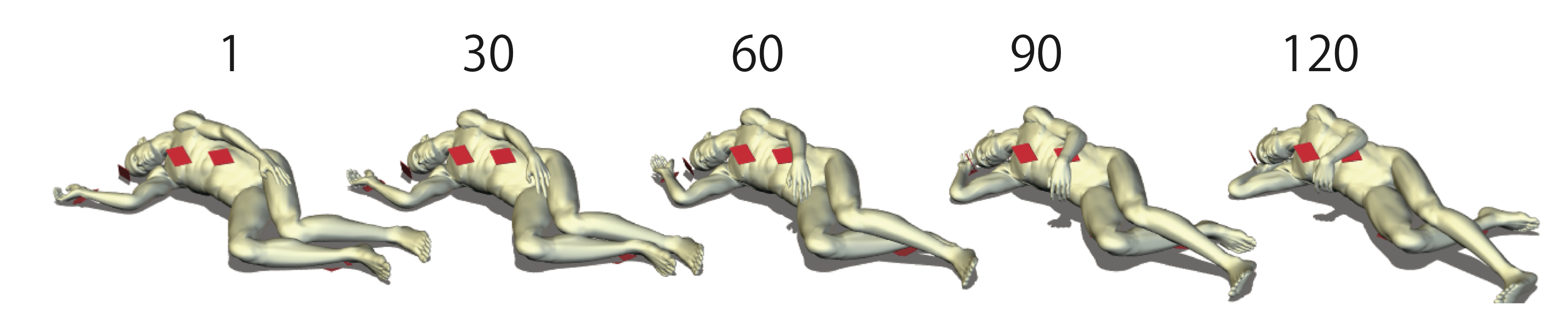}
 \caption{Sleeping}
 \end{subfigure}
 \begin{subfigure}{0.9\columnwidth}
 \centering
 \includegraphics[width=\textwidth,height=2.5cm]{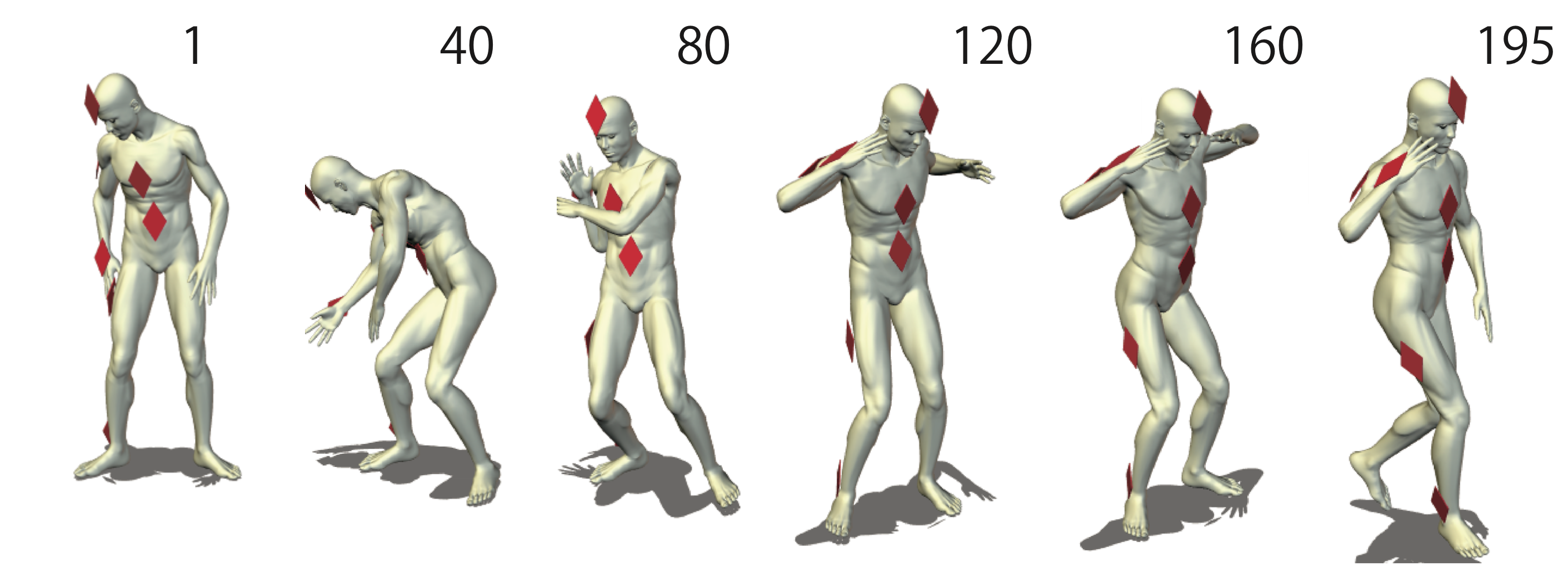}
 \caption{Wearing a jacket}
 \end{subfigure}
 \caption{Postures used in~\cite{channel} to model the WBAN channel (Pictures source: \cite{channel})}
 \label{fig:postures}
\end{figure}

The nodes positions are calculated in $7$ postures: walking (walk), walking weakly (weak), running (run), sitting down (sit), wearing a jacket (wear), sleeping (sleep), and lying down (lie). these postures are represented on Fig. \ref{fig:postures}. Walk, weak, and run are variations of walking motions. Sit and lie are variations of up-and-down movement. Wear and sleep are relatively irregular postures and movements.
\newline \indent Channel attenuation is calculated between each couple of nodes for each of these positions as the average attenuation (in dB) and the standard deviation (in dBm). The model takes into account: the shadowing, reflection, diffraction, and scattering by body parts.

\subsection{Simulation Settings}

Above the channel model described in the previous section, we used standard protocol implementations provided by the Mixim framework. In particular, we used, for the medium access control layer, the IEEE 802.15.4 implementation. The sensitivity levels, header length of the packets and other basic information and parameters are taken from the 802.15.4 standards.

Each data point is the average of 50 simulations run with different seeds. The simulations are performed with only one packet transmission from the gateway (i.e the sink). The transmission power is set at the minimum limit level $-60\,dBm$ that allows an intermittent communication given the channel attenuation and the receiver sensitivity $-100\,dBm$. With this value of transmission power, we guarantee that at each time t of the simulation, we have a connected network and at the same time we ensure a limited energy consumption.

\subsection{Simulation results}
To compare between the different broadcast strategies presented in the previous section \ref{BS} (\footnote{Note that we performed an additional study on the timer used in the MBP strategy in order to choose the one offering the best compromise. The results of this study are reported in \ref{MBP-timer}}), we studied for each one three basic parameters:
\begin{itemize}
 \item \textbf{Percentage of covered nodes:} Since our unique source is the sink which broadcasts a single message in the network, we therefore calculate the percentage of nodes that have received this message.
 \item \textbf{End-to-end delay}:The average End-to-end delay is the time a message takes to reach the destination(i.e. every node except the sink).
 \item\textbf{Number of transmissions and receptions}:This is a key parameter, it helps us to have an estimation of the energy consumption at each node and for each posture for all the strategies.
\end{itemize}

\subsubsection{Percentage of covered nodes}

\begin{figure}[htbp]
\centering
\includegraphics[width=0.9\columnwidth]{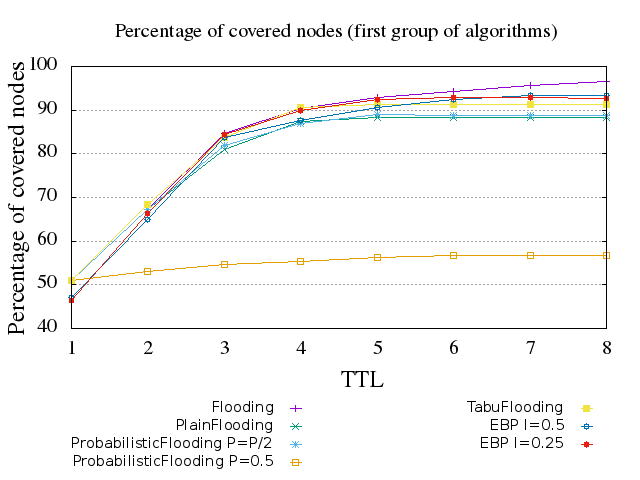}
\caption{Percentage of covered nodes (first group)}
\label{Fig2}
\end{figure}

\begin{figure}[htbp]
\centering
\includegraphics[width=0.9\columnwidth]{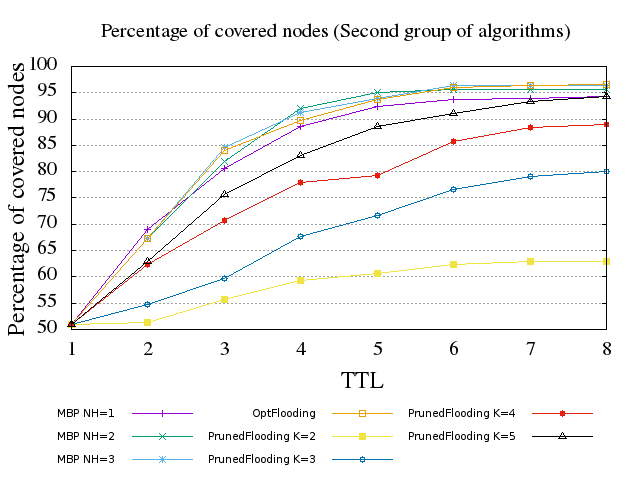}
\caption{Percentage of covered nodes (Second group)}
\label{Fig3}
\end{figure}

\begin{figure}[htbp]
\centering
\includegraphics[width=0.9\columnwidth]{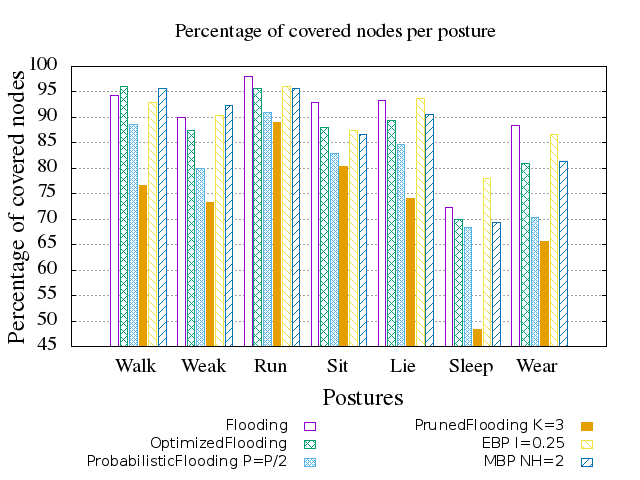}
\caption{Percentage of covered nodes per posture}
\label{Fig4}
\end{figure}

We first compare the percentage of covered nodes after the broadcast of the message from the sink. Figures \ref{Fig2} and \ref{Fig3} show this percentage variations for each algorithm in function of the \emph{TTL} "Time-To-Live" set for the packets. The choice of the \emph{TTL} is justified by the fact that it is the only parameter in common between all the considered algorithms.
\newline \indent The first observation is about the value of the percentage of covered nodes when \emph{TTL} is equal to one. As shown in figures \ref{Fig2} and \ref{Fig3}, it is almost the same for all strategies and is equal to 51\%\ except for \emph{EBP}. This value shows that half of the network is covered after the first transmission by the sink. For these simulations, human body posture is \emph{walk posture}. In this posture, the sink has at least three neighbors that can reach at the first time which explain the percentage of covered nodes with $TTL=1$ (the other postures are shown in the figure \ref{Fig4}). The slight decrease seen with \emph{EBP} strategy is due to packet losses caused by collisions with control packets.

The strategy \emph{Flooding} represents the reference of flooding-based strategies. With this strategy nodes continue to retransmit the packet while its \emph{TTL} is greater than $1$, regardless of the past. For this reason, and in absence of conditions on retransmissions, strategy \emph{Flooding} shows the highest percentage of covered nodes.
For \emph{EBP}, $I$ represents the interval between two successive control message \emph{Hello message}. Each node send periodically a \emph{Hello message} to discover the neighborhood since the decision to rebroadcast or not the message depends on the neighborhood state. \emph{EBP} shows good results with a slight increase when $I=0.25$. Indeed, with $I=0.5$, the node has a less realistic view of the neighborhood which could force it to decide to not rebroadcast the message.
\newline \indent A notable observation is that EBP stops automatically the broadcast when the each node's neighbors are covered, while the classical \emph{Flooding} algorithm (Figure \ref{Fig2}) has to wait until the \emph{TTL} of all packets reaches $1$.
\newline \indent \emph{Plain Flooding} and \emph{Tabu Flooding} coverage starts to increase with the \emph{TTL} and stabilize rapidly. Both algorithms stop even when the network is not fully covered. In case of \emph{Plain Flooding} a few number of transmissions is allowed, only once, which corresponds to the first reception of the message. For \emph{Tabu Flooding}, nodes only rebroadcast messages, if it is intended for them, and only for uncovered node. So in case the destination is not in the sender's range, the message will be dropped by the other nodes. As it is very hard for these two algorithms to reach a full coverage, the notions of completion delay and the required amount of transmissions to cover the whole network are meaningless. Therefore these algorithms will be left out of the subsequent evaluations.
\newline \indent The \emph{Probabilistic Flooding} with decreasing probability ($P=P/2$) achieves a better coverage than the constant probability version with $P=0.5$. In fact, with \emph{Probabilistic Flooding}, $P$ initially equal to $1$, which means that the probability to rebroadcast a message for the first time is equal to $1$ too, since the chosen random variable is necessarily less than $P$.

For \emph{Pruned Flooding} strategy , $K$ represents the number of nodes, a node chooses randomly to forwards the message to. In order to study the influence of this parameter on the behavior of the strategy, we varied the value of $K$ from $2$ to $5$. Figure\ref{Fig3} shows that the number of nodes to choose randomly has a direct effect on the coverage percentage. With $K=2 and K=3$ it is almost impossible to cover all the nodes even with higher values of \emph{TTL}. When $K=5$, we have better results but still the lowest percentage comparing to the other strategies.
\newline \indent For \emph{MBP}, $NH$ represents the threshold on the total number of hops the message passed through. We therefore studied the influence of this parameter on \emph{MBP} results. Figure\ref{Fig3} shows a slight variation depending on $NH$ parameter especially between $NH=2$ and $NH=3$. In general, \emph{MBP} shows a good percentage of covered nodes,specially with low values of \emph{TTL}, for example, with $TTL=3$, the strategy covers between $80$\%\ and $85$\%\ (with different values of $NH$) of the nodes.
\newline \indent The novel strategy \emph{Optimized Flooding} has good results. For example, with $TTL=4$, $90$\%\ of the network is covered. As this strategy is an adaptation of the strategy \emph{Flooding}, the percentage of covered nodes values and even the variation in function of the TTL are almost the same.

Figure \ref{Fig4} represents the network coverage percentage in different postures.
\newline \indent The lowest percentage, independently of the posture, is with \emph{Pruned Flooding} strategy, with $K=3$ a low number of nodes are designated to rebroadcast the message, knowing that a node doesn't rebroadcast a message destined to another node.
\newline \indent \emph{Probabilistic Flooding} shows a better percentage than \emph{Pruned Flooding} strategy but still the lowest.
Both strategies have higher variations among the postures due to their random nature.
\newline \indent Strategy \emph{Flooding} shows the highest percentage in almost all postures. However, \emph{MBP}, \emph{EBP} and \emph{Optimized Flooding} show good percentage of covered nodes between $70$\%\ and $96$\%. For these strategies and for the percentage of covered nodes parameter, even if a variation in function of the postures is observed, it still less significant than for the two other studied parameters: End-To-End delay (\ref{EED}) and number of transmissions and receptions (\ref{TXRX}). It is important to point that the sleep posture is a specific case that represents a static posture where some nodes are hidden and remain so all along the simulation. Even so, \emph{Optimized Flooding} reaches $70$\%\ of covered nodes.

\subsubsection{Average End-To-End Delay}
\label{EED}

\begin{figure}[htbp]
\centering
\includegraphics[width=0.9\columnwidth]{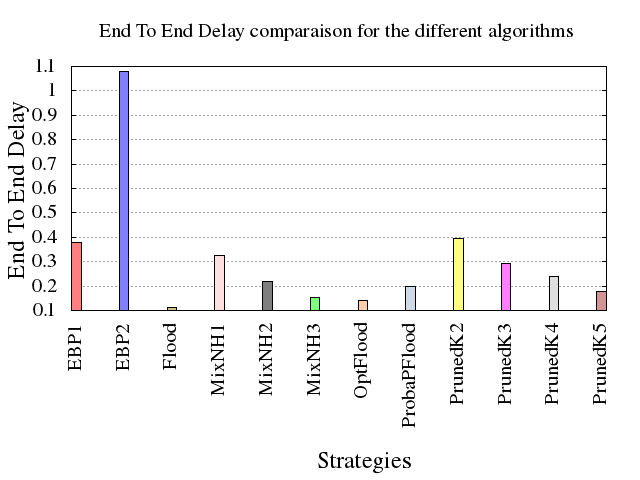}
\caption{Average End-to-End delay comparison between strategies}
\label{Fig55}
\end{figure}

Figure \ref{Fig55} represents the average End-to-End delay, i.e. the average time required for a message sent from the sink to reach a node.
\newline \indent As expected, \emph{Flooding} algorithm has the best performance due to the huge amount of message copies in the network.
\newline\indent The strategy \emph{Probabilistic Flooding (P=P/2)} shows a higher end-to-end delay due to the decreasing amount of retransmissions with respect to the flooding.
\newline \indent The end-to-end delay achieved by \emph{EBP} is strictly dependent from the \emph{Hello message} interval ($I$). When $I$ is short, for instance $I=0.25s$, each time the connections in the model change, it is soon considered by nodes, resulting in a precise acquisition of the network state and a quick decision making.
\newline \indent \emph{Pruned Flooding} end-to-end delay is also directly related to the value of $K$ parameter: the higher is the value of $K$, the lower is the end-to-end delay. In particular in cases when $K=2$ and $K=3$, the algorithm spends a lot of time trying to cover the whole network. This is due to the low number of nodes that receive the packet and are designated to retransmit.
\newline \indent \emph{MBP} presents better performances in term of End-to-End delay than \emph{EBP} and \emph{Pruned Flooding}, even if, with this strategy, a node delays packet retransmission after T timer expiration. \emph{MBP} End-to-End delay depends also on $NH$ value. With $NH=3$, \emph{MBP} attempts to cover the nodes more quickly than for lower values, because, in this case, nodes use to rebroadcast the message faster based on simple flooding without delaying the retransmission.
\newline \indent Finally, \emph{Optimized Flooding} has an end-to-end delay close to \emph{Flooding} strategy and it is also the best End-To-End delay after \emph{Flooding}. Our first goal is reached since the idea is to maintain the low End-To-End delay given by \emph{Flooding}.

\begin{figure}[htbp]
\centering
\includegraphics[width=0.9\columnwidth]{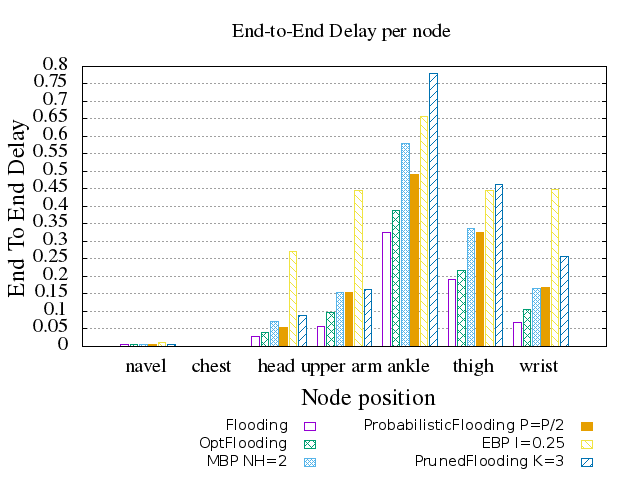}
\caption{Average End-to-End Delay per node}
\label{Fig6}
\end{figure}

Figure \ref{Fig6} details the average End-to-End delay per node.
\newline \indent We can notice that all algorithms behave similarly: they spend more time to cover the peripheral nodes in the network (example: node at the ankle). Indeed, peripheral nodes are highly mobile and located at a greater distance from the sink than the other ones. As a result, even if an algorithm is able to cover the central part of the network in a very short time, it then has to wait for a connection opportunity with the last nodes.
\newline \indent \emph{EBP} and \emph{pruned Flooding} present the worst End-To-End delay. In case of \emph{EBP}, nodes wait before retransmitting the message until all conditions are satisfied, i.e, the number of nodes in the neighborhood.
\newline \indent Both of \emph{Flooding} and \emph{Optimized Flooding} strategies have the lowest End-To-End delay for all the nodes. These results are especially important regarding the end-to-end delay of the peripheral nodes. \emph{Optimized Flooding} comes to deal moderately with this limit comparing to the other strategies including \emph{MBP}.

\begin{figure}[htbp]
\centering
\includegraphics[width=0.9\columnwidth]{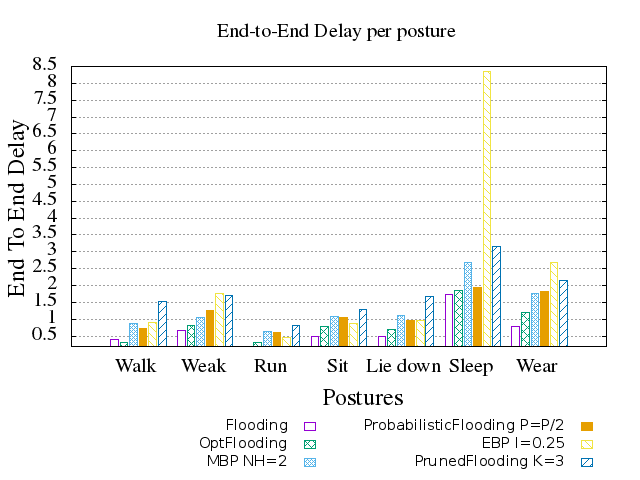}
\caption{Average End-to-End delay per posture}
\label{Fig7}
\end{figure}

Figure \ref{Fig7} shows the average End-To-End delay per posture.
\newline \indent \emph{EBP} seems to perform better in the high mobility postures. Indeed, in \emph{EBP}, because the number of neighbors is important, a peak is noticed in sleep position where some nodes are hidden and remain so all along the simulation, for that, nodes decide to not rebroadcast the message and keep it until all conditions are satisfied.
\newline \indent In other hand, \emph{Flooding}, \emph{Optimized Flooding} and \emph{MBP} are the less affected by human body postures than \emph{EBP} and \emph{Pruned Flooding}. However, we notice better results (closest to \emph{Flooding}) and less variations in function of the postures with \emph{Optimized Flooding} than with \emph{MBP}.

\subsubsection{Number of transmissions and receptions}
\label{TXRX}

The number of transmissions and receptions is a key parameter for WBAN network. The number of transmissions reflects the channel load and also gives an indication about the amount of electro-magnetic energy that will be absorbed by the body. The number of receptions gives an indication on the energy consumption of the different nodes in the network and hence on the devices autonomy, their capability to rely on a reduced size battery or even to harvest energy.
\begin{figure}[htbp]
\centering
\includegraphics[width=0.9\columnwidth]{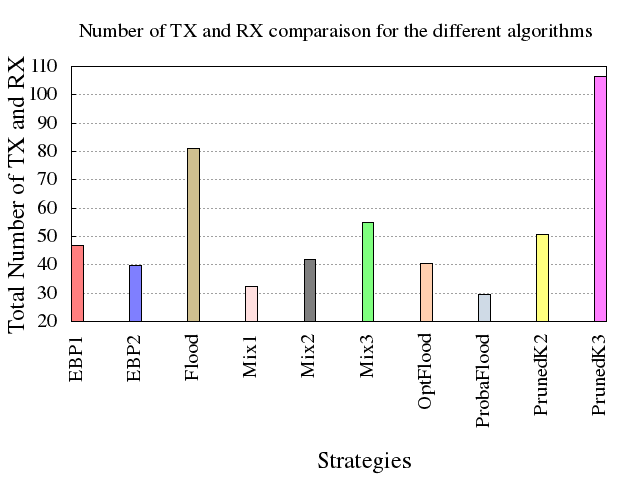}
\caption{Number of transmissions and receptions}
\label{Fig8}
\end{figure}

Figure \ref{Fig8} compares the total number of transmissions and receptions for all the studied algorithms.
\newline \indent \emph{Flooding} pays its good End-To-End delay and coverage performance as it exhibits the highest number which indicates a high energy consumption. It is the only algorithm that has no limitation on the number of transmissions besides the \emph{TTL}.
\newline \indent The number of transmissions of \emph{EBP} varies with $I$. In fact, \emph{EBP} strongly relies on the transmission of \emph{hello messages} that cannot be neglected, this causes collisions with data packets (especially with shorter interval $I$) leading to lots of dropped packets and to unnecessary retransmissions. However it is important to note here that these results only account for data packets transmissions and receptions.
\newline \indent The adaptive \emph{Probabilistic Flooding (P=P/2)} has a better performance than the classical \emph{Flooding}, because the probability decrease limits the number of retransmissions as the packet travels in the network.
\newline \indent The energy consumed by \emph{Pruned Flooding} depends on the parameter $K$: while for the end-to-end delay and coverage, the algorithm performs well with high values of $K$, in this case a visible reduction in number of transmissions and receptions is observed with lower values of $K$. For each retransmission, the node duplicates the packet as much as the number of nodes to choose randomly i.e as much as $K$ value.
\newline \indent \emph{MBP} does not suffer from the same issue, even if control messages such as acknowledgments are also necessary. In fact, acknowledgments are only sent when a packet is effectively forwarded, the control traffic is directly related to the network activity. \emph{MBP} performance therefore mostly depends on the $NH$ parameter and lowering this value results in a better energy efficiency.
\newline \indent Finally, with the strategy \emph{Optimized Flooding}, a significant decrease in the number of transmissions and receptions is observed: $50$\%\ compared to \emph{Flooding}. It's an important result. Remember that the unique limitation of \emph{Flooding} strategy is related to the number of transmissions and receptions in other words the energy consumptions.

\begin{figure}[htbp]
\centering
\includegraphics[width=0.9\columnwidth]{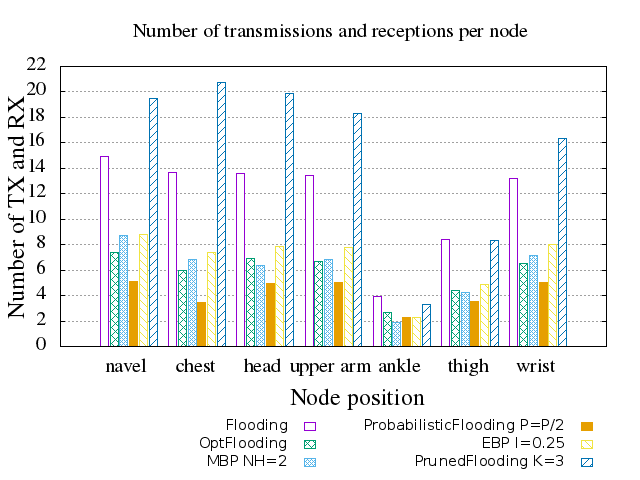}
\caption{Number of transmissions and receptions per node}
\label{Fig9}
\end{figure}

Figure \ref{Fig9} shows the total number of transmissions and receptions for each node.
\newline \indent We observe that nodes at the center of the network (navel and chest) have the highest number of transmissions and receptions, their position allows them to communicate with more nodes. On the contrary the node on the ankle is the one with the lowest number of transmissions and receptions because standing in the periphery has few occasions to communicate with the rest of the network which explain again the highest End-To-End delay for this node. Besides, when the message reaches this node (node at the ankle), its \emph{TTL} has already been decremented multiple times and it generally has no uncovered neighbor as far as knowledge-based algorithms are concerned.
\newline \indent \emph{Flooding} and \emph{Pruned Flooding} present the highest number for all nodes. Nodes continue to rebroadcast messages while $TTL$ is greater than $1$. It is worst with \emph{Pruned Flooding} where each message is duplicated three times (these results are for $K=3$).
\newline \indent \emph{Probabilistic Flooding (P=P/2)} performs better than \emph{Flooding} and \emph{Pruned Flooding}. Indeed, the central nodes happen to transmit more than the others, but their neighbors will limit their transmission probability immediately, resulting in a lower number of receptions. This algorithm therefore better distributes the consumption across the network.
\newline \indent One of the main weaknesses of the flooding algorithms that their performance strongly depends on the \emph{TTL}. Increasing the \emph{TTL} lets the total amount of transmissions and receptions rise very quickly. This issue is partially solved in the adaptive \emph{Probabilistic Flooding (P=P/2)}: the probability decrease works as an automatic brake, drastically reducing the transmission probability.
\newline \indent For \emph{MBP} and \emph{EBP} the total number of transmissions and receptions is $50$\%\ less than \emph{Flooding} and \emph{Pruned Flooding}. Indeed, \emph{MBP} uses the information from the acknowledgement to control and stop the transmissions, while \emph{EBP} uses information included in the control messages (\emph{Hello message}).
\newline \indent Even if, for our strategy \emph{Optimized Flooding}, no control message or acknowledgement are exchanged between nodes, we observe a significant decrease on the total number of transmissions and receptions per node. The decision to stop the retransmission is related to the validity of the information contained in each copy of the message. This information is used to stop sending the message if the network is fully covered as well as eliminate the back and forth of the same copy between two nodes, especially between highly connected nodes.

\begin{figure}[htbp]
\centering
\includegraphics[width=0.9\columnwidth]{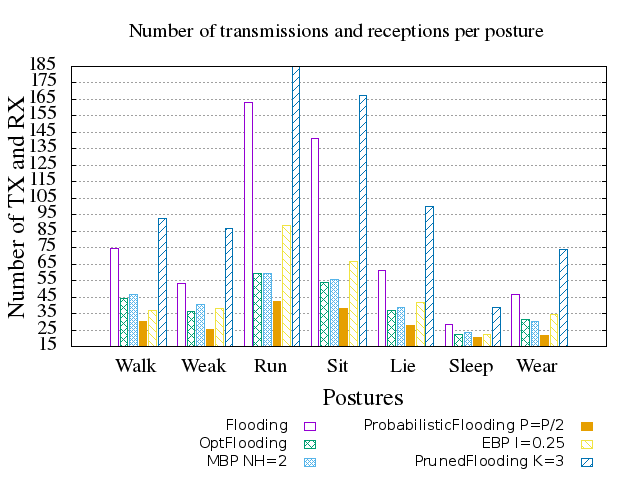}
\caption{Number of transmissions and receptions per posture}
\label{Fig11}
\end{figure}

Figure \ref{Fig11} represents the number of transmissions and receptions for each posture. It's important to study each strategy behavior in function of the wearer posture.
\newline \indent We can notice, on this graph, for all strategies, the highest number is for \emph{Run and Sit} postures. In fact, \emph{Run} posture represents the highest mobility, even with frequent disconnections, nodes are able to meet more frequent to exchange messages. In \emph{Sit} posture, nodes are closer to each other so they are able to exchange packets with a minimum loss.
\newline \indent \emph{Sleep} posture presents the lowest number of transmissions and receptions for all strategies. 
At a first glance, it seems an advantage for this posture but reconsidering the observations related to the percentage of covered nodes \ref{Fig4} and to the end-to-end delay \ref{Fig7}, this is rather related to the high disconnections between nodes.
\newline \indent The two strategies \emph{MBP} and \emph{Optimized Flooding} are less independent from the postures (ie less variations). Still, \emph{Optimized Flooding} presents a lower number of transmissions and receptions than \emph{MBP} except for some postures which could be explained by the fact that \emph{Optimized Flooding} covers more nodes as shown in figure \ref{Fig4}.


\section{Total order resilience}
In this section we push further our study in order to detect the capacity of the strategies described to be reliable and to ensure the total order message delivery.
That is, messages sent by the sink in a specific order should be received by each node in the system in that specific order.

We stress strategies with transmission rate ranging from $1$ to $1000$ packets per second broadcasted from the sink to the other nodes. We fixed the MAC buffer capacity to 100\footnote{Additional simulations for various MAC buffer capacities are presented in Section \ref{impact-mac}.}.

We study two parameters: 
\begin{enumerate}
\item Percentage of covered nodes: we calculate the number of messages received at each node than we deduce the percentage.
\item Percentage of desequencing: the percentage of messages received in the wrong order.
\end{enumerate}

\paragraph{Percentage of covered Nodes}

 \begin{figure}[htbp]
 \begin{subfigure}{0.9\columnwidth}
 \centering
 \includegraphics[width=\textwidth,height=3.25cm]{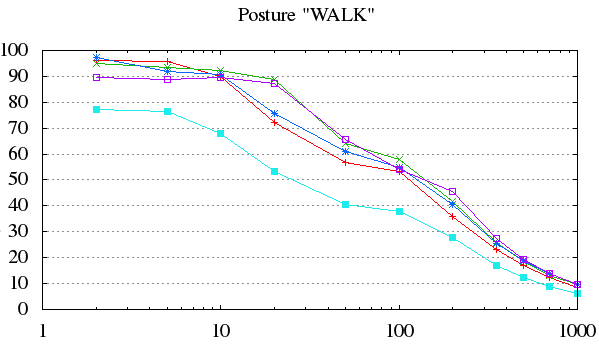}
 \end{subfigure}
 \begin{subfigure}{0.9\columnwidth}
 \centering
 \includegraphics[width=\textwidth,height=3.25cm]{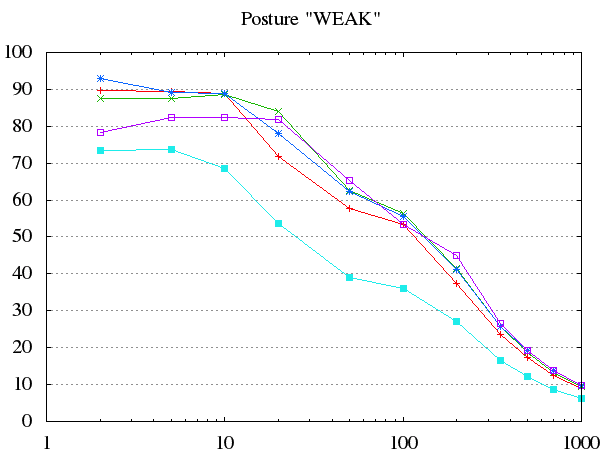}
 \end{subfigure}
 \begin{subfigure}{0.9\columnwidth}
 \centering
 \includegraphics[width=\textwidth,height=3.25cm]{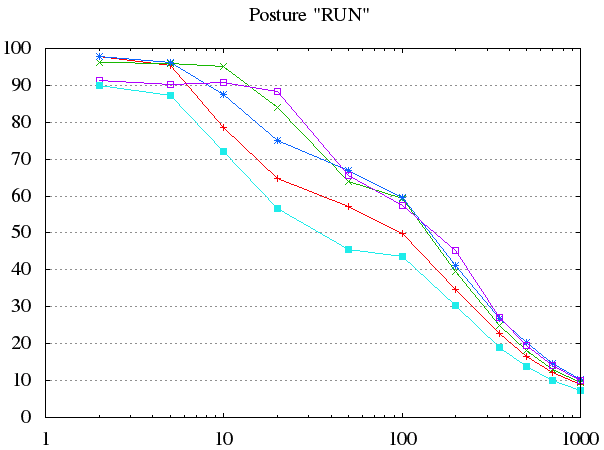}
 \end{subfigure}
 \begin{subfigure}{0.9\columnwidth}
 \centering
 \includegraphics[width=\textwidth,height=3.25cm]{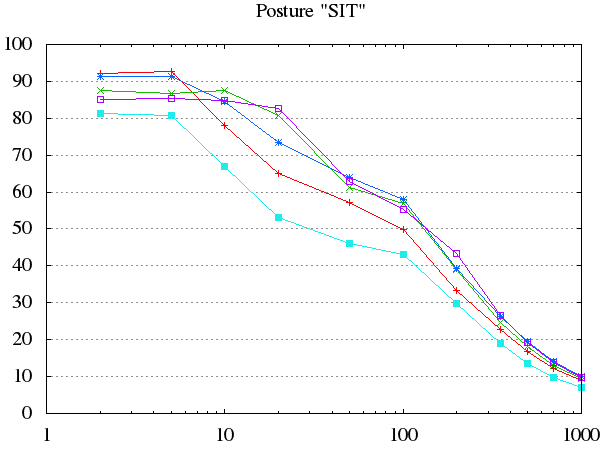}
 \end{subfigure}
 \begin{subfigure}{0.9\columnwidth}
 \centering
 \includegraphics[width=\textwidth,height=3.25cm]{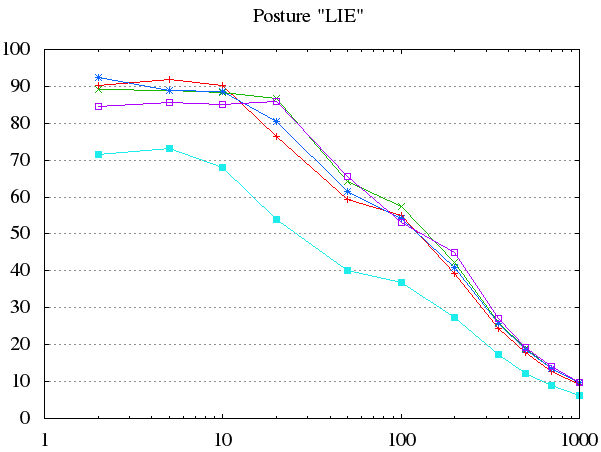}
 \end{subfigure}
 \begin{subfigure}{0.9\columnwidth}
 \centering
 \includegraphics[width=\textwidth,height=3.25cm]{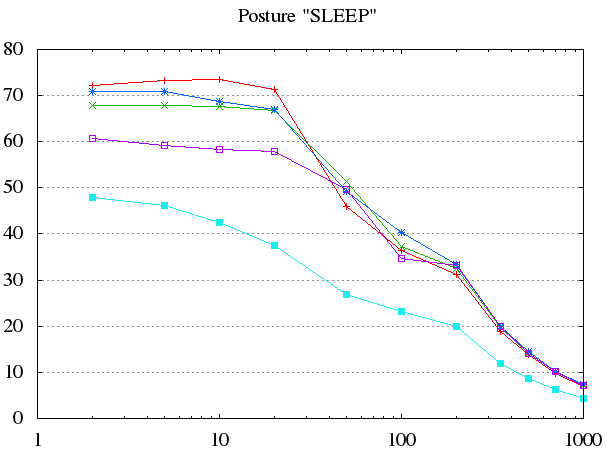}
 \end{subfigure}
 \begin{subfigure}{0.9\columnwidth}
 \centering
 \includegraphics[width=\textwidth,height=4.75cm]{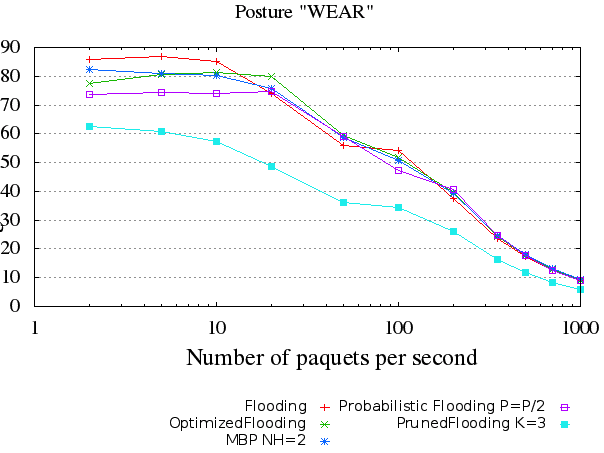}
 \end{subfigure}
 \caption{Percentage of covered nodes per posture}
 \label{ReceivedMsg1}
\end{figure}

Figure \ref{ReceivedMsg1} presents the percentage of covered nodes in function of packets number per second.

Contrary to the expectations, all studied strategies behave similarly: when the packets rate goes to $1000$ packets per second the percentage of covered nodes almost linearly decreases to $10$\%. At $100$ packets per second, percentage of covered nodes barely exceeds $50$\%.

For \emph{Flooding} and \emph{MBP} percentage stagnates until $10$ packets per second than the curve starts decrease. However, For \emph{Optimized Flooding} and \emph{Probabilistic Flooding}, percentage remains constant until $20$ packets/s

\emph{PrunedFlooding} presents the lowest percentage. Other strategies results are quite close at the beginning and starting from certain point curves overlap and converge to the same point.

\paragraph{Percentage of Desequencing}

\begin{figure}[htbp]
 \begin{subfigure}{0.9\columnwidth}
 \centering
 \includegraphics[width=\textwidth,height=3.25cm]{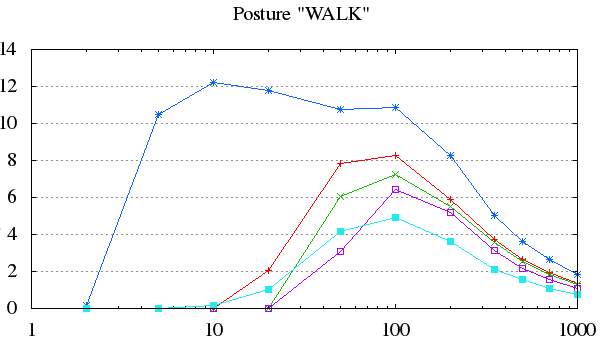}

 \end{subfigure}
 \begin{subfigure}{0.9\columnwidth}
 \centering
 \includegraphics[width=\textwidth,height=3.25cm]{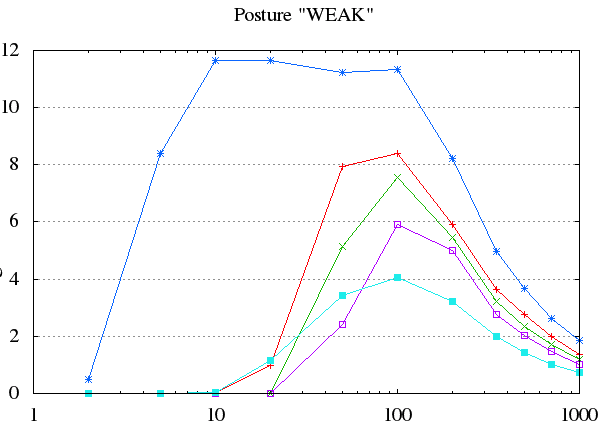}

 \end{subfigure}
 \begin{subfigure}{0.9\columnwidth}
 \centering
 \includegraphics[width=\textwidth,height=3.25cm]{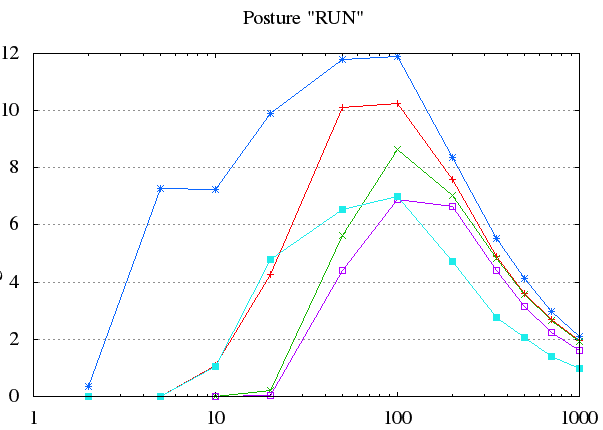}

 \end{subfigure}
 \begin{subfigure}{0.9\columnwidth}
 \centering
 \includegraphics[width=\textwidth,height=3.25cm]{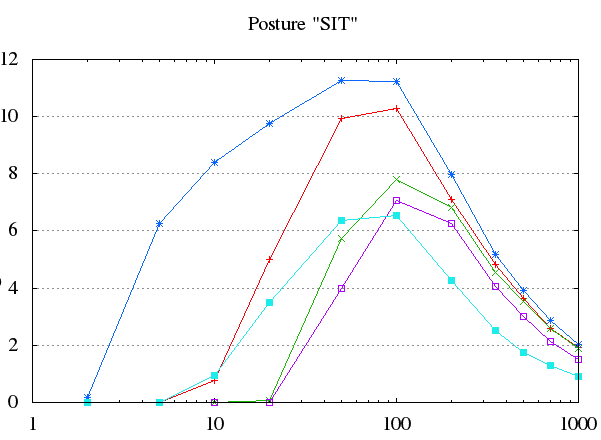}

 \end{subfigure}
 \begin{subfigure}{0.9\columnwidth}
 \centering
 \includegraphics[width=\textwidth,height=3.25cm]{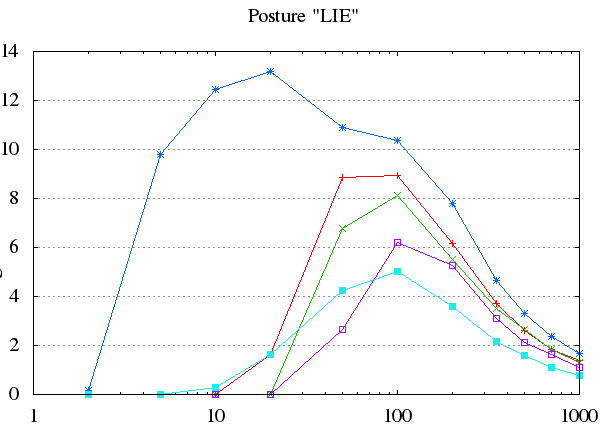}

 \end{subfigure}
 \begin{subfigure}{0.9\columnwidth}
 \centering
 \includegraphics[width=\textwidth,height=3.25cm]{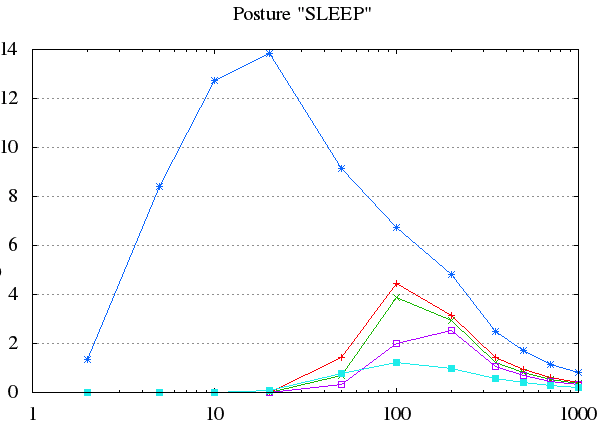}

 \end{subfigure}
 \begin{subfigure}{0.9\columnwidth}
 \centering
 \includegraphics[width=\textwidth,height=4.75cm]{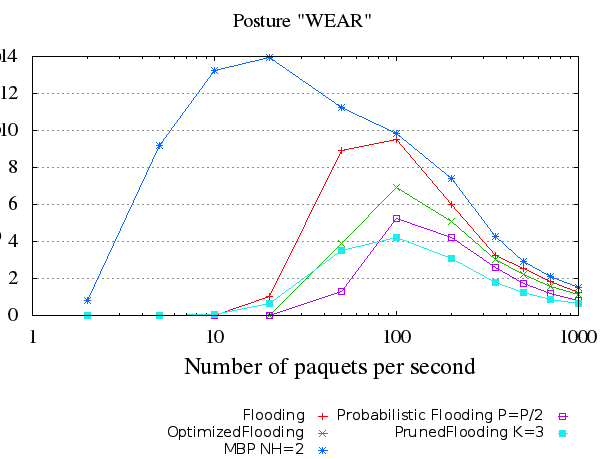}
 \end{subfigure}

 \caption{Percentage of Desequencing per posture}
 \label{Deseq1}
\end{figure}

Figure \ref{Deseq1} presents the percentage of desequencing.

Contrary to results about the percentage of covered nodes, from curves, three phases can be seen:
\begin{itemize}
\item At the beginning, all strategies present $0$\% of desequencing. At this point, strategies are able to handle more than one packet in the network.
\item Then, from certain point (depending on the strategy), this percentage increases linearly. Here, referring to Figure \ref{ReceivedMsg1}, percentage of covered nodes decreases due to collisions and packets loss. Sequencing is no longer ensured.
\item Finally, from certain point, percentage decreases to converge to 0\% again. Also, referring to figure \ref{ReceivedMsg1}, this is due to the fact that few packets are received.
\end{itemize}

\emph{MBP} strategy presents the highest percentage of desequencing starting from $2$ packets/s.
 
Percentage of desequencing increases starting from $10$ packets/s for \emph{Flooding} and \emph{Pruned Flooding}, and from $20$ packets/s for \emph{Optimized Flooding} and \emph{Probabilistic Flooding}.

\emph{PrunedFlooding} low percentage of desequencing is due to a low percentage of covered nodes.

\emph{Optimized Flooding} strategy presents the lowest percentage of desequencing compared to \emph{Flooding} and \emph{MBP} strategies.





\section{Conclusion and future works}
In this paper we evaluated through simulation the performance of several DTN-inspired broadcast strategies in a WBAN context. Our simulations, realized with the Omnet++ simulator, the Mixim framework and a WBAN channel model proposed in the literature and issued from realistic data-sets, allowed us to compare flooding-like strategies that forward packets blindly and differ mostly by how their stopping criterion with a representative knowledge-based algorithm, EBP, which relies on the knowledge of the neighborhood of each node and its evolutions. The simulations realized over a 7 nodes network in 7 types of movements allow a fine characterization of the compromise that exists between the capacity to flood the whole network quickly and the cost induced by this performance. Simulations also allowed us to identify some less intuitive behaviors: for all strategies, most of the time is spent trying to reach leaf nodes, which makes us think that the key lies in adaptive algorithms that are able to mix different strategies.

We described a novel protocols that relies on such an adaptive approach: \emph{MBP}, for Mixed Broadcast Protocol that applies a more aggressive strategy in the center of the network, where connections are more stable, and becomes more cautious at the border of the network, where a blind transmission has a good chance of success. This protocol has been reported in the extended abstract of this work \cite{BCPP15}.
Additionally we proposed the \emph{Optimized Flooding} protocol that improves the performances of \emph{MBP}. Furthermore, we propose a preliminary study related to the total order reliability of the existing strategies. This study advocates for further investigations on this specific direction.

As future works, we will adapt this work to the context of IR-UWB transmissions, which is a promising technology for WBAN. Another future work would be a detailed study of existing channel models \cite{CHModel1} and a comparison with the channel model. Furthermore, another interesting future direction would be to consider collisions among multiple WBANs \cite{WBAN3,WBAN2}.

\section{Appendix: Additional simulations}

\subsection{Impact of MAC queue length}
\label{impact-mac}
\begin{figure}[htbp]
 \centering
 \includegraphics[width=0.9\columnwidth]{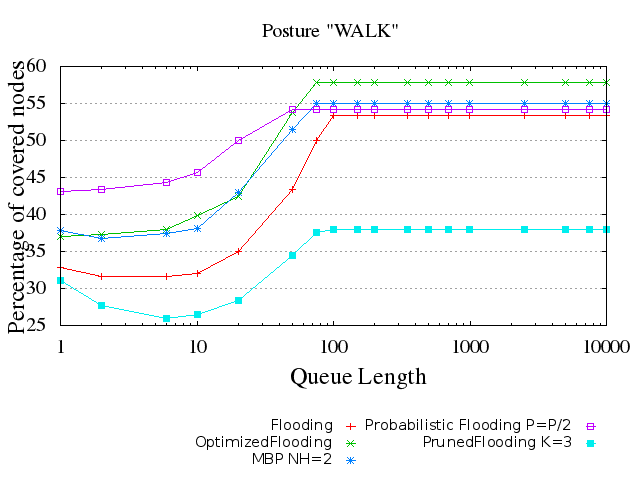}
 \caption{Percentage of covered nodes for different queue length }
 \label{ReceivedMsgMAC1}
\end{figure}

 \begin{figure}[htbp]
 \centering
 \includegraphics[width=0.9\columnwidth]{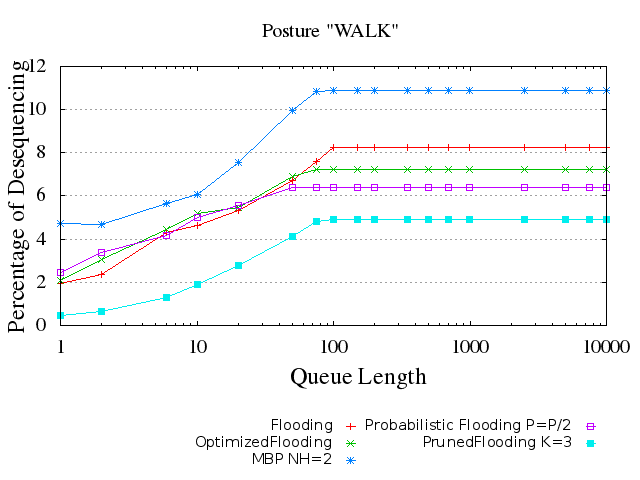}
 \caption{Percentage of Desequencing for different queue length}
 \label{DeseqMAC1}
\end{figure}

We run simulations with $100$ packets per second and we vary MAC buffer capacity from $1$ to $10000$ packets.
Figures \ref{ReceivedMsgMAC1} and \ref{DeseqMAC1} show respectively the percentage of covered nodes and the percentage of desequencing function of the MAC queue length.
\emph{Flooding} shows less resilience to high transmission rate.
\emph{Probabilistic Flooding} is less affected by queue length in MAC layer.
Our novel strategy stabilizes to the highest 
highest percentage of covered nodes.
\emph{MBP} shows again a high percentage of desequencing.

We can conclude that MAC queue length influences our strategies performance while its value is lower than the transmission rate. For a queue length superior than the transmission rate, no variation is observed in the percentage of covered nodes nor in the percentage of desequencing. 

\begin{figure}[htbp]
 \centering
 \includegraphics[width=0.9\columnwidth]{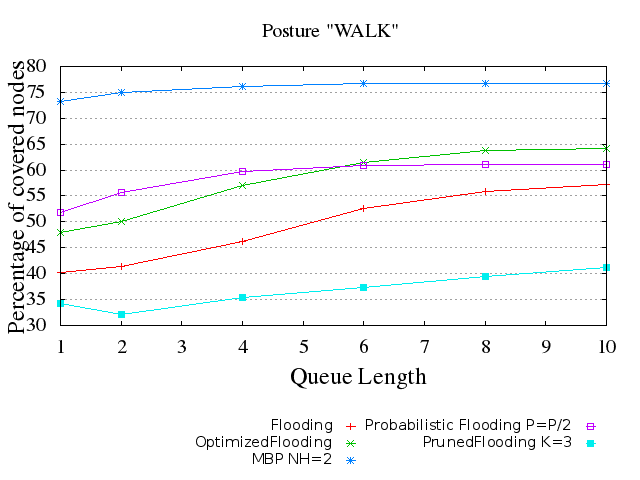}
 \caption{Percentage of covered nodes for different queue length }
 \label{ReceivedMsgMAC2}
\end{figure}

\begin{figure}[htbp]
 \centering
 \includegraphics[width=0.9\columnwidth]{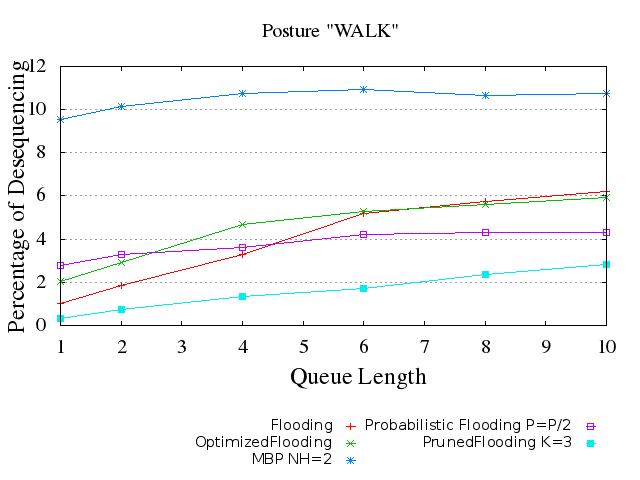}
 \caption{Percentage of Desequencing for different queue length}
 \label{DeseqMAC2}
\end{figure}

In the following we fixe the transmission rate at 10 packets per second (the rate usually used in the biomedical applications) and a variation of the queue length from 1 to 10. The simulations results are proposed in Figures \ref{DeseqMAC2} and \ref{ReceivedMsgMAC2}.

\subsection{MBP: Timer study}
\label{MBP-timer}
In the following we focus on the timer used for \emph{MBP} strategy. The objective of this study is to have a closer inside on the \emph{MBP} parameters (i.e. timer) impact on the overall performances of this strategy.Recall that in \emph{MBP} strategy, starting from a certain threshold on the number of hops ($NH$), a node delays message broadcasting after a timer expiration. Thus, we will examine the influence of this timer on \emph{MBP} strategy performances.
%
%
For simulations, we considered $NH$ equal to $2$. We also varied timer's value to obtain $15$ different values from $0.005$ to $1.000$ second.

\begin{figure}[htbp]
\centering
 \begin{subfigure}{0.8\columnwidth}
 \centering
 \includegraphics[width=\textwidth,height=3.7cm]{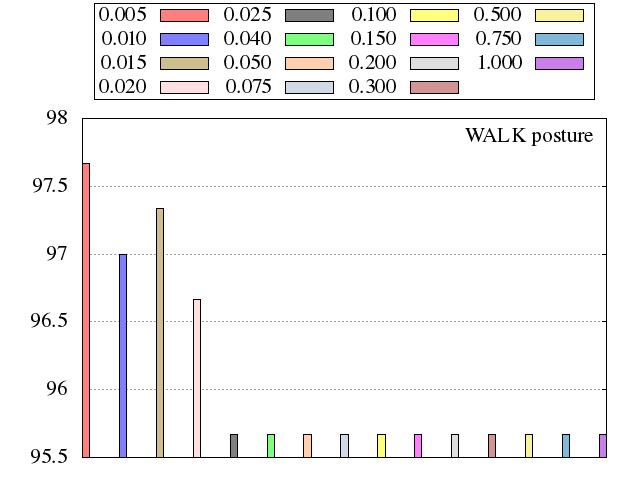}
 \end{subfigure}
 \begin{subfigure}{0.8\columnwidth}
 \centering
 \includegraphics[width=\textwidth,height=3cm]{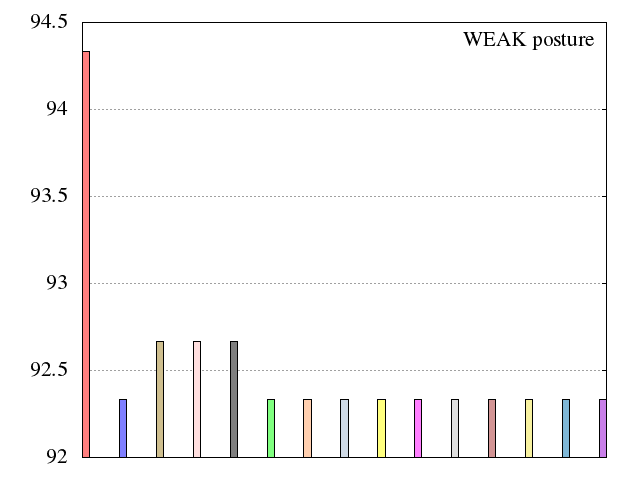}
 \end{subfigure}
 \begin{subfigure}{0.8\columnwidth}
 \centering
 \includegraphics[width=\textwidth,height=3cm]{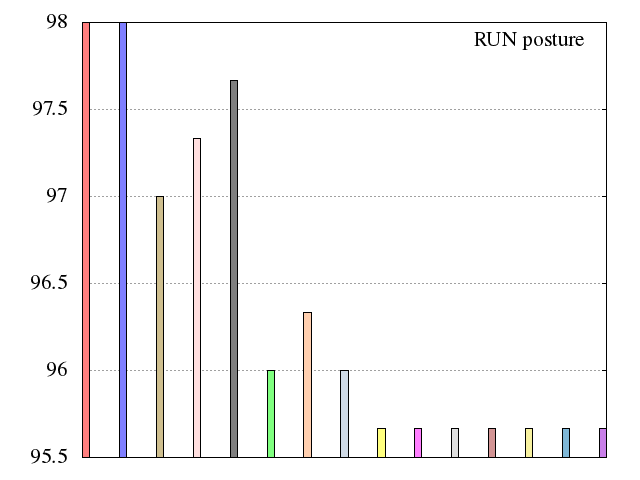}
 \end{subfigure}
 \begin{subfigure}{0.8\columnwidth}
 \centering
 \includegraphics[width=\textwidth,height=3cm]{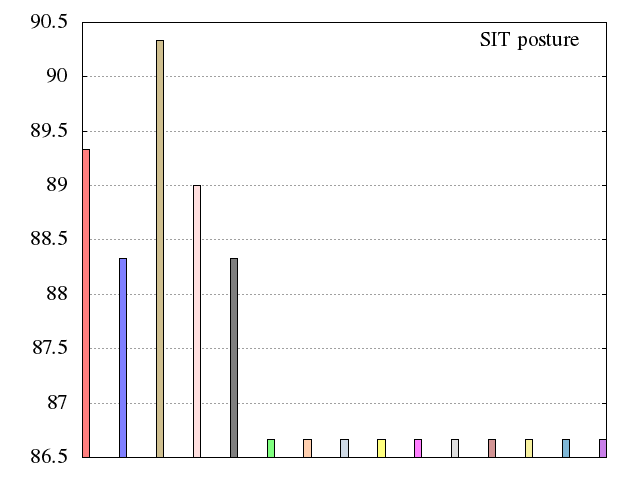}
 \end{subfigure}
 \begin{subfigure}{0.8\columnwidth}
 \centering
 \includegraphics[width=\textwidth,height=3cm]{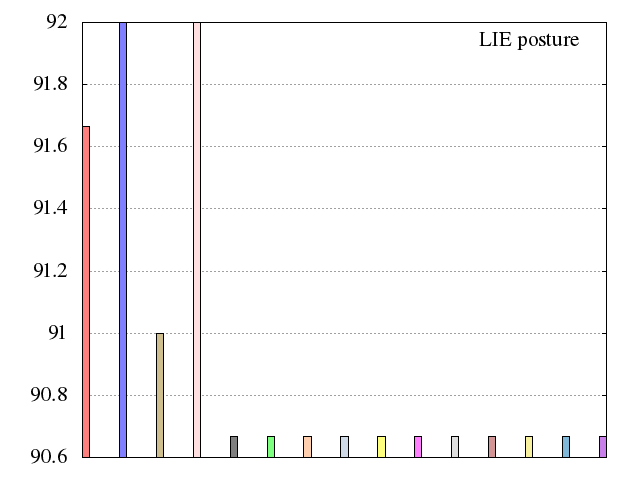}
 \end{subfigure}
 \begin{subfigure}{0.8\columnwidth}
 \centering
 \includegraphics[width=\textwidth,height=3cm]{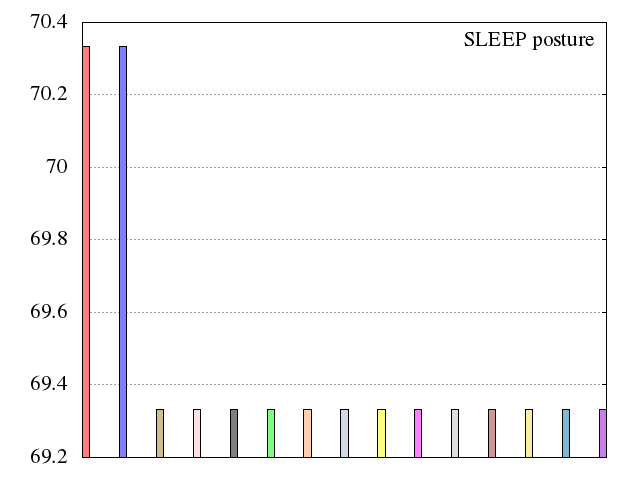}
 \end{subfigure}
 \begin{subfigure}{0.8\columnwidth}
 \centering
 \includegraphics[width=\textwidth,height=3cm]{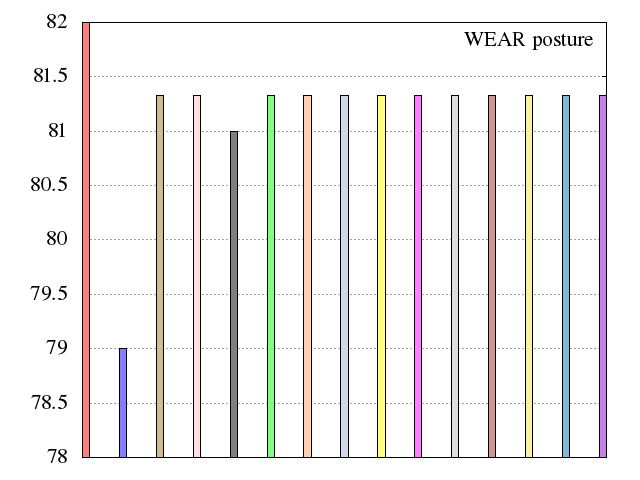}
 \end{subfigure}
 \caption{Percentage of covered nodes per posture for each timer}
 \label{CoveredNodesMBPTimer}
\end{figure}

Figure \ref{CoveredNodesMBPTimer} shows the percentage of covered nodes for each timer value per posture.

For all postures, the obtained percentages for each timer value are extremely close: the average difference is around $2.5$\%.

We also noted that with the increase of the timer value, the simulation results are closer to each other.

The first four timer values (the smallest ones) show better percentage for covered nodes. Short timer pushes nodes to broadcast more often the message in the network which increases the probability of receiving the message by the neighboring nodes.

\begin{figure}[htbp]
\centering
 \begin{subfigure}{0.8\columnwidth}
 \centering
 \includegraphics[width=\textwidth,height=3.7cm]{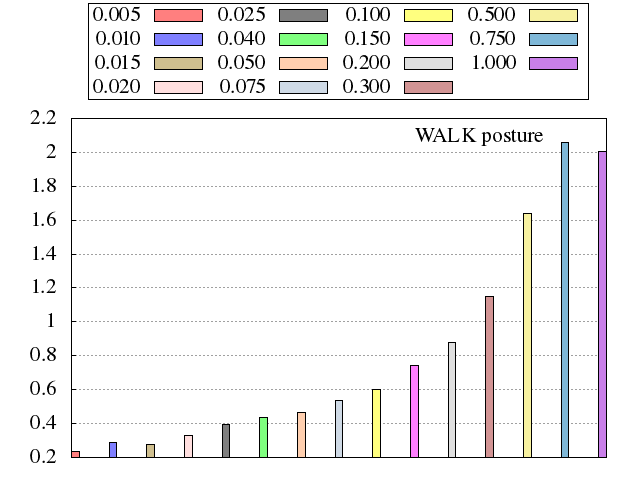}
 \end{subfigure}
 \begin{subfigure}{0.8\columnwidth}
 \centering
 \includegraphics[width=\textwidth,height=3cm]{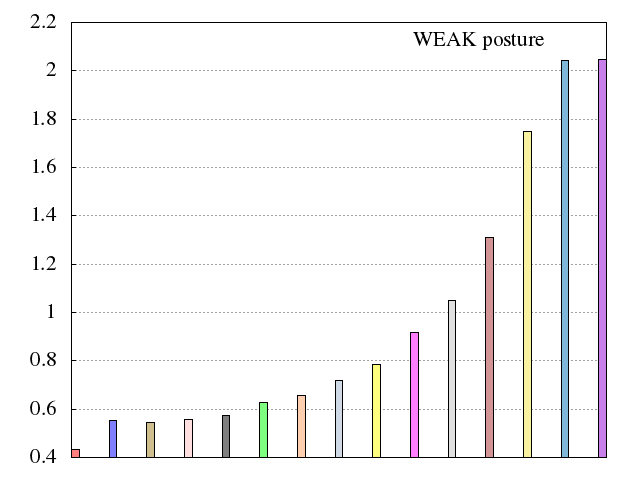}
 \end{subfigure}
 \begin{subfigure}{0.8\columnwidth}
 \centering
 \includegraphics[width=\textwidth,height=3cm]{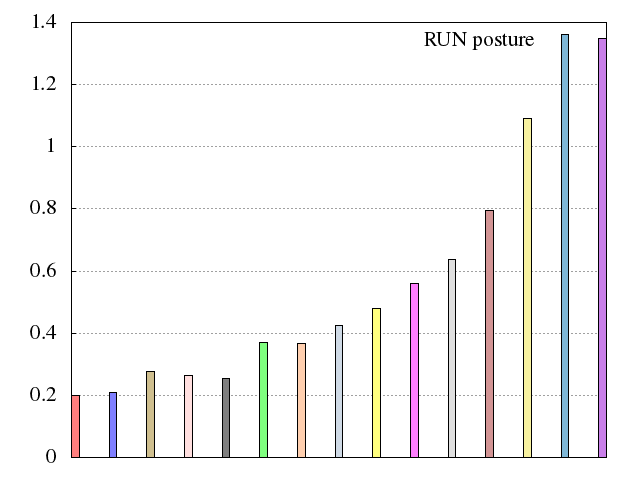}
 \end{subfigure}
 \begin{subfigure}{0.8\columnwidth}
 \centering
 \includegraphics[width=\textwidth,height=3cm]{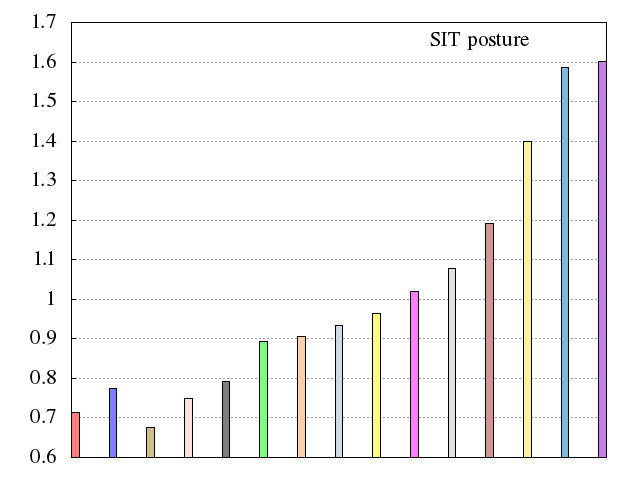}
 \end{subfigure}
 \begin{subfigure}{0.8\columnwidth}
 \centering
 \includegraphics[width=\textwidth,height=3cm]{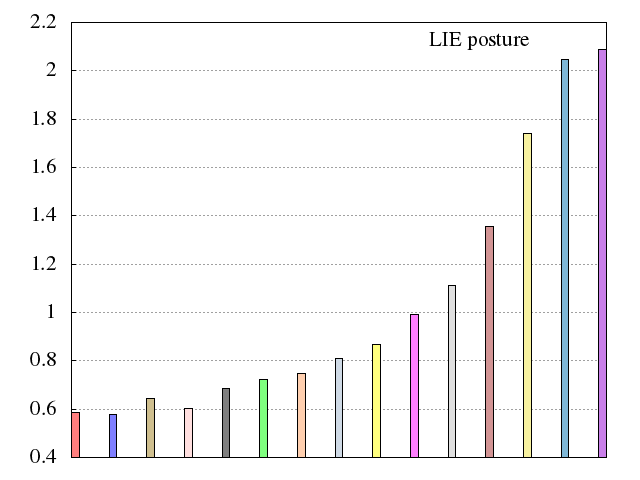}
 \end{subfigure}
 \begin{subfigure}{0.8\columnwidth}
 \centering
 \includegraphics[width=\textwidth,height=3cm]{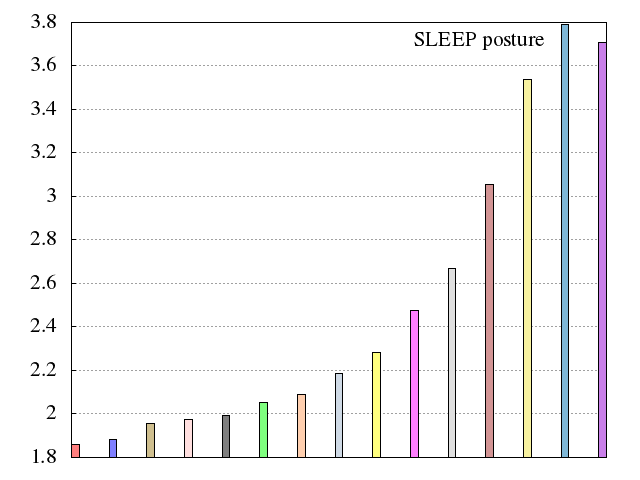}
 \end{subfigure}
 \begin{subfigure}{0.8\columnwidth}
 \centering
 \includegraphics[width=\textwidth,height=3cm]{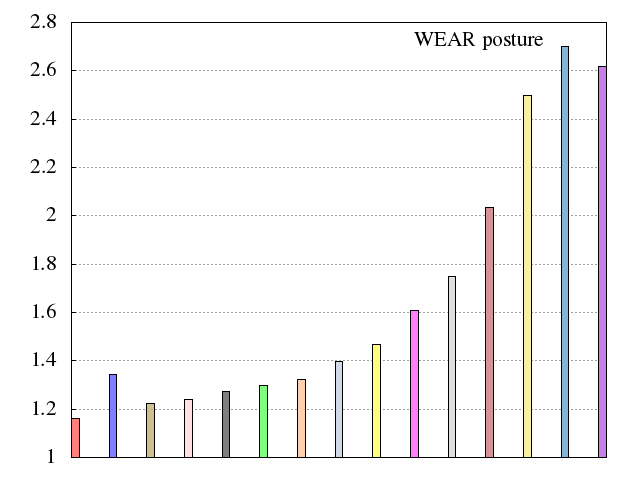}
 \end{subfigure}
 \caption{Average End-To-End Delay per posture for each timer}
 \label{EndToEndDelayMBPTimer}
\end{figure}

Figure \ref{EndToEndDelayMBPTimer} shows the average end-to-end delay for each timer value per posture.
%
%
It should be noted the performance of MBP strategy. With the increase of the timer value, the MBP strategy spends more time to cover all nodes in the network.
The best end-to-end delay is obtained with timer's value equal to $0.005$ second.

\begin{figure}[htbp]
\centering
 \begin{subfigure}{0.8\columnwidth}
 \centering
 \includegraphics[width=\textwidth,height=3.7cm]{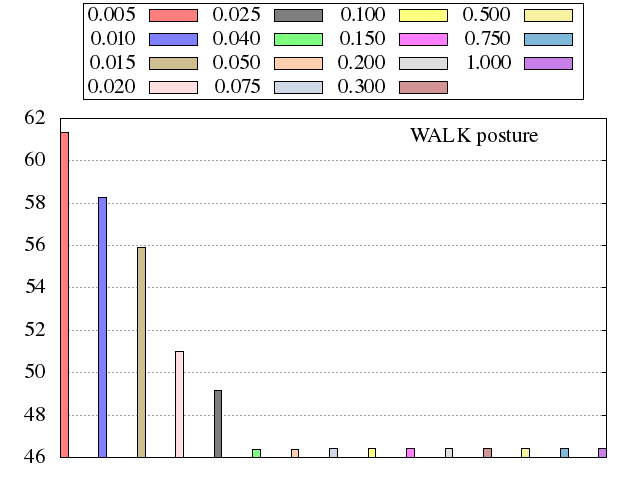}
 \end{subfigure}
 \begin{subfigure}{0.8\columnwidth}
 \centering
 \includegraphics[width=\textwidth,height=3cm]{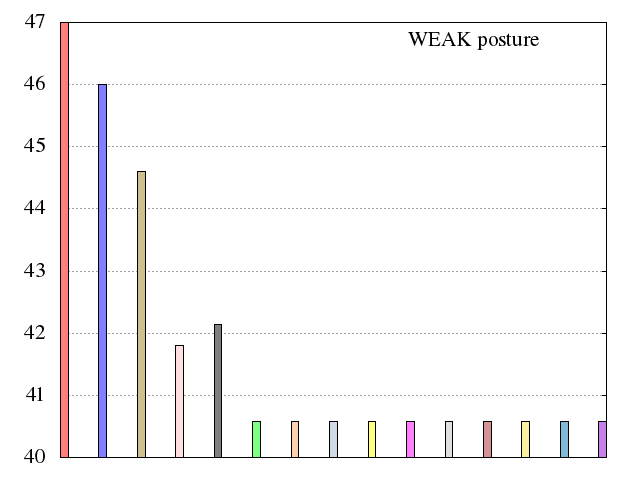}
 \end{subfigure}
 \begin{subfigure}{0.8\columnwidth}
 \centering
 \includegraphics[width=\textwidth,height=3cm]{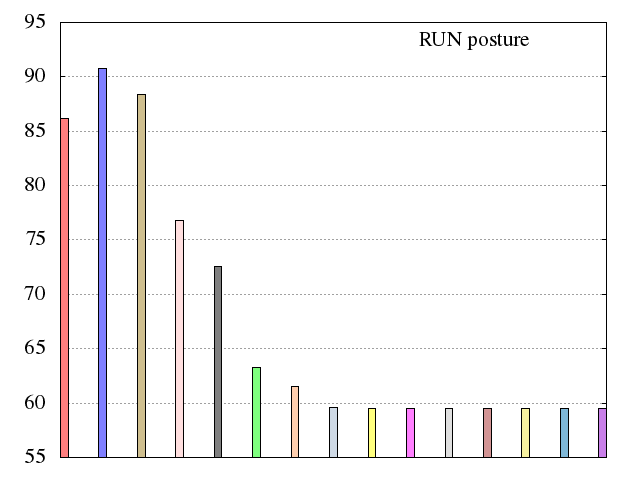}
 \end{subfigure}
 \begin{subfigure}{0.8\columnwidth}
 \centering
 \includegraphics[width=\textwidth,height=3cm]{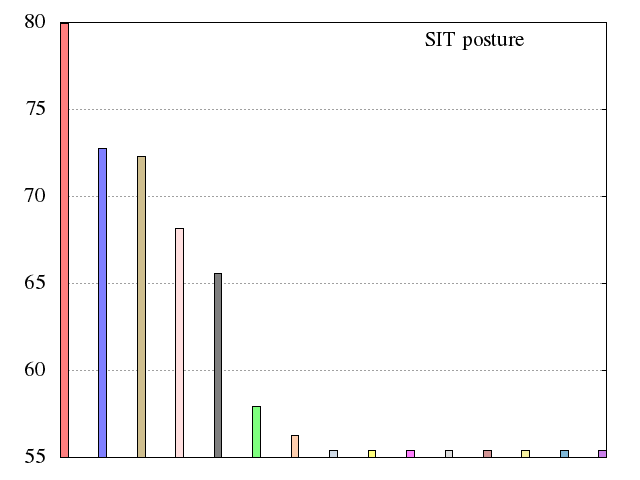}
 \end{subfigure}
 \begin{subfigure}{0.8\columnwidth}
 \centering
 \includegraphics[width=\textwidth,height=3cm]{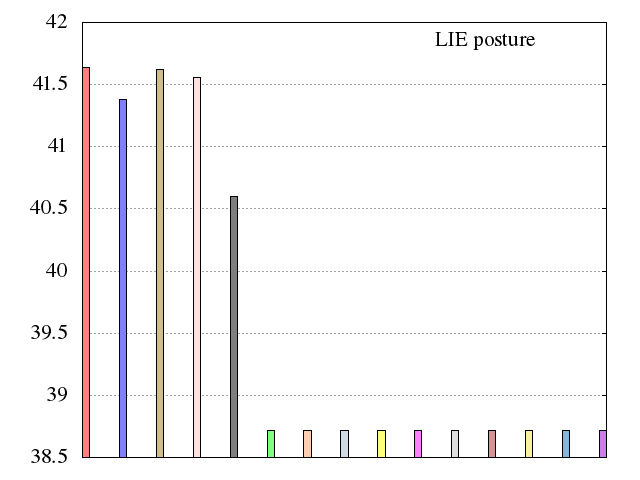}
 \end{subfigure}
 \begin{subfigure}{0.8\columnwidth}
 \centering
 \includegraphics[width=\textwidth,height=3cm]{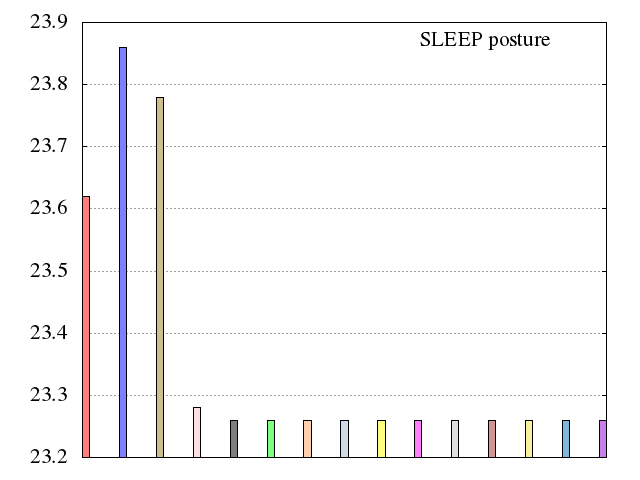}
 \end{subfigure}
 \begin{subfigure}{0.8\columnwidth}
 \centering
 \includegraphics[width=\textwidth,height=3cm]{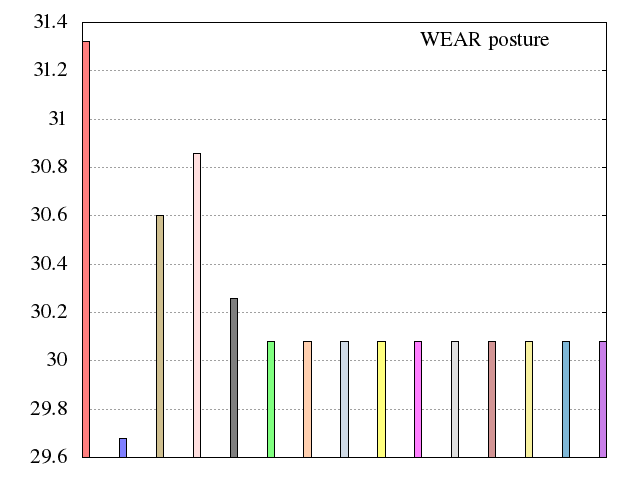}
 \end{subfigure}
 \caption{Number of transmissions and receptions per posture for each timer}
 \label{nbreTxRxMBPTimer}
\end{figure}

Figure \ref{nbreTxRxMBPTimer} shows the number of transmissions and receptions for each timer per posture.
Contrary to the end-to-end delay parameter, we obtain a descending curve. This is due to the fact that when the timer is longer, nodes spend more time waiting before rebroadcasting the message. Obviously, this has a direct impact on the average end-to-end delay.
However, short timers impact the energy consumption (see Figure \ref{nbreTxRxMBPTimer}).


%
%

%
%
%
%
%
%
%
%
%
\bibliographystyle{abbrv}
\bibliography{biblioJournal}

\end{document}